\documentclass[authoryear,letterpaper,twocolumn,5p]{elsarticle}
\usepackage[T1]{fontenc}
\usepackage[latin1]{inputenc}
\usepackage{graphicx}
\usepackage{amssymb}
\usepackage{amsmath}
\usepackage{wasysym}
\usepackage{natbib}
\usepackage{lineno}
\usepackage[pdftex,colorlinks]{hyperref}

\makeatletter 
\newcommand\thedate{\@date}
\makeatother

\makeatletter
\def\ps@pprintTitle{%
     \let\@oddhead\@empty
     \let\@evenhead\@empty
     \def\@oddfoot{\footnotesize\itshape
 	Preprint deposited on arxiv.org.
       \hfill\thedate}%
}
\makeatother

\begin{document}
\begin{frontmatter}
\title{Global and Local Sea Level During the Last Interglacial:\\ A Probabilistic Assessment
}

\author[geo,wws]{Robert E. Kopp\corref{cor1}} \ead{rkopp@princeton.edu}
\author[geo]{Frederik J. Simons}
\author[geo]{Adam C. Maloof}
\author[geo,wws]{Michael Oppenheimer}

\cortext[cor1]{Corresponding author}

\address[geo]{Department of Geosciences and} \address[wws]{Woodrow Wilson School of Public and International Affairs, \\Princeton University, Princeton, NJ 08544, USA}

\date{March 2, 2009}

\begin{abstract}

The Last Interglacial (LIG) stage (ca. 130--115~ka), with polar temperatures likely 3--5$^\circ$C warmer than today, serves as a partial analogue for low-end future warming scenarios. Multiple indicators suggest that LIG global sea level (GSL) was higher than at present; based upon a small set of local sea level indicators, the Intergovernmental Panel on Climate Change (IPCC)'s Fourth Assessment Report inferred an elevation of approximately 4--6 m.  While this estimate may be correct, it is based upon overly simplistic assumptions about the relationship between local sea level and global sea level. Sea level is often viewed as a simple function of changing global ice volume. This perspective neglects local variability, which arises from several factors, including the distortion of the geoid and the elastic and isostatic deformation of the solid Earth by shifting ice masses. Accurate reconstruction of past global and local sea levels, as well as ice sheet volumes, therefore requires integrating globally distributed data sets of local sea level indicators. To assess the robustness of the IPCC's global estimate and search for patterns in local sea level that are diagnostic of meltwater sources, we have compiled a comprehensive database that includes a variety of local sea level indicators from 47 localities, as well as a global sea level record derived from oxygen isotopes. We generate a global synthesis from these data using a novel statistical approach that couples Gaussian process regression to Markov Chain Monte Carlo simulation of geochronological errors. Our analysis strongly supports the hypothesis that global sea level during the Last Interglacial was higher than today, probably peaking between 6--9 m above the present level. This level is close to that expected from the complete melting of the Greenland Ice Sheet, or from major melting of both the Greenland and West Antarctic Ice Sheets. In the period when sea level was within 10 m of the modern value, the fastest rate of sea level rise sustained for a 1 ky period was likely about 80--110 cm per century. Combined with the evidence for mildly higher temperatures during the LIG, our results highlight the vulnerability of ice sheets to even relatively low levels of sustained global warming.
\end{abstract}

\begin{keyword}
Sea level change, Pleistocene, data analysis
\end{keyword}

\end{frontmatter}



\makeatletter
\def\@oddhead{\small{\textsc{Sea Level During the Last Interglacial\hfill Kopp et al.}}}
\makeatother

\section{Introduction}

As a result of industrial activity, greenhouse gas concentrations now exceed levels reached on Earth at any time within the last eight hundred thousand years \citep{Solomon2007a}. The long-lived greenhouse gases in the atmosphere today produce a radiative forcing of about 2.6 W/m$^2$. Given a climate sensitivity of 2--4.5$^\circ$C per doubling of carbon dioxide levels \citep{Meehl2007a}, the equilibrium global warming expected from this forcing -- without considering the effects of any further increases in greenhouse gas concentrations -- is between 1.4--3.2$^\circ$C. Among the many effects expected to accompany this warming, global sea level rise, driven primarily by thermal expansion of seawater and melting ice sheets, features prominently \citep{Meehl2007a}. Uncertainties in ice sheet behavior make it difficult to predict sea level rise precisely, but by the end of the twenty-first century, mean global sea level rise could exceed one meter \citep[e.g., ][]{Rahmstorf2007a,Grinsted2009a}. Since changes of this magnitude have no precedent in human experience, to understand them and to compile observations against which to test models of future climate change, it is necessary to turn to the geological record.

Data from past interglacial stages and earlier warm periods may help constrain the future behavior of sea level. Synthesizing geological sea level indicators into a global reconstruction requires taking into account interpretative uncertainties, geochronological uncertainties, and regional variability. Differences between local and global sea level arise from several factors, including the distortion of the geoid and the elastic and isostatic deformation of the solid Earth by shifting ice masses \citep{Farrell1976a,Mitrovica2003a}. Accurate reconstruction of past global and local sea levels therefore requires integrating global data sets of local sea level indicators.

In this paper, we develop a novel statistical approach to assessing the probability distribution of local and global sea level through time from a database of geographically distributed sea level indicators. Our algorithm is based upon Gaussian process regression with a Markov Chain Monte Carlo treatment of geochronological errors. We apply this method to a new database of geographically dispersed sea level indicators spanning the Last Interglacial (LIG) stage, which climaxed about 125 thousand years ago.
The LIG (also known as the Eemian stage, its local northern European name, and as Marine Isotope Stage 5e) is of special interest for three reasons: (1) it is recent enough that it may be possible to construct relatively high-resolution sea level records for this interval, (2) due in large part to enhanced northern hemisphere insolation, mean global temperature may have been slightly warmer than at present,  and (3) several lines of evidence suggest global sea level was higher than today, perhaps by 4--6 m \citep{Jansen2007a}.

Greenhouse gas concentrations during the LIG were comparable to pre-Industrial Holocene levels \citep{Petit1999a}, but Earth's orbital eccentricity during the Last Interglacial was 0.038--0.041, more than twice as high as the modern value \citep{Berger1991a}. Energy balance modeling predicts that, as a consequence, summer temperatures between 132--124~ka  on all land masses except Antarctica were at least 0.5$^\circ$C warmer than today \citep{Crowley1994a}, while a more complete climate model indicates summer temperatures 2--4$^\circ$C warmer than today in most of the Arctic \citep{Otto-Bliesner2006a}. Ice core data from both Greenland and Antarctica suggest polar temperature anomalies in both hemispheres of about 3--5$^\circ$C \citep{Jansen2007a}, comparable to the 3--6$^\circ$C of Arctic warming expected to accompany 1--2$^\circ$C of global warming \citep{Katsov2004a}. In Europe, pollen data suggest middle Eemian summer temperatures about 2$^\circ$C warmer than present \citep{Kaspar2005a}. While the global mean temperature anomaly is uncertain, sea-surface temperatures in the equatorial Pacific \citep{Lea2004a} and Atlantic \citep{Weldeab2007a} were about 2$^\circ$C warmer than pre-Industrial levels. 

For our analysis of LIG sea level, we compiled an extensive database of sea level records from this time period. Sea level can be inferred from two basic sources: the oxygen isotopic composition of seawater as preserved in carbonates, and geomorphological and sedimentological records of local sea level. The oxygen isotopic composition of seawater is in large part a function of ice sheet volume and is thus closely related to global sea level. The latter is defined as the integrated local sea level over the entire ocean, which is a measure of ocean volume. Water enriched in the light isotope $^{16}$O is concentrated preferentially in the ice sheets; therefore, as ice sheets melt and global sea level rises, the relative concentration of the heavy isotope, $\delta^{18}$O, of seawater decreases. The record of marine oxygen isotopes is thus a proxy for global sea level, although its interpretation is complicated by several factors. The oxygen isotopic composition of marine water is also a function of local salinity, while the difference between the isotopic composition of water and the carbonates precipitated from it is a function of temperature \citep{Emiliani1966a}. The interpretation of the carbonate oxygen isotope record is further complicated by the histories of different water masses, by changes in their distribution over time, and, on long time scales, by isotopic exchange with the mantle \citep{Jean-Baptiste1997a,Schrag2002b}. Nonetheless, with assumptions about the isotopic composition of ice sheets and a paleotemperature constraint, it is possible to estimate ice volume and thus sea level from $\delta^{18}$O in carbonate, as we do later.

Many variables affect the local relative sea level recorded in sedimentary facies and in geomorphology. Aside from ocean volume, these factors include the effect of the distribution of ice, water, and sediment on the geoid, lithospheric flexure, isostatic adjustments, the orientation and magnitude of Earth's spin axis, and shoreline geometry \citep{Farrell1976a,Mitrovica2003a}, as well as tectonic uplift and thermal subsidence. As a result of gravitational and elastic effects, local sea level shortly after an ice sheet melting event falls in the near-field and rises by more than the global average in the far-field. Over the following several millennia, isostasy slowly emplaces a mass of mantle to compensate for the missing ice, thereby changing the geoid so that local sea level in both the far-field and the near-field relax toward the global average. (In the near-field, however, the vertical motion of the solid Earth complicates the observed relative sea level change.) For this reason, local sea levels at Pacific islands in the far-field of Laurentide Ice Sheet were 1--3 m higher in the middle Holocene than today \citep{Mitrovica2002a}. Thus, even if LIG ice volume never shrunk below present levels and mean global sea level never exceeded its present value, local sea levels several meters higher than present could have occurred in the far-field of the Laurentide Ice Sheet early in the LIG, and comparably high local sea levels could have occurred in its intermediate-field late in the LIG \citep{Lambeck1992a}. Without accurate and precise dating of the relevant sea level indicators, these effects could produce the false appearance of a magnified global sea level high-stand. In order to estimate ice sheet history from sea level records, it is  thus necessary to account for these factors. Conversely, because many of these effects lead relative sea-level changes to differ in the near- and far-fields of an ice sheet, a global database of local sea level indicators could address not just whether ice volume was smaller during the Last Interglacial than today, but also what combination of melting ice sheets, if any, was responsible for higher global sea levels.

Our goal is to use as comprehensive a database as possible to produce probability distributions of sea level as a function of geographic location and geological age. We have to cope with variable temporal uncertainty, as well as with variable errors in sea level measurements. In addition, some of the data are censored in that they provide only an upper or lower bound to sea level. Where possible, we also want to take advantage of quasi-continuous sequences, in which relative timing is known with greater precision than absolute dates. This is the case for the global oxygen isotope curve from benthic foraminifera \citep{Lisiecki2005a}, as well as for sequences of local observations of sedimentary facies from the Netherlands \citep{Zagwijn1983a} and of sea levels derived from hydrological modeling of foraminiferal oxygen isotopes from the Red Sea \citep{Rohling2008a}.

Our statistical approach is based upon Gaussian process regression \citep{Rasmussen2006a}. This method, of which the commonly-used geospatial technique of kriging interpolation is a well-known example,  treats a field (such as sea level) as a collection of random variables drawn from a Gaussian distribution.  By specifying the covariance structure of the field,  knowledge about the relevant physics affecting the process can be incorporated into the modeling without constraining it to fit a particular forward model. With a sufficiently precise and accurate data set, such an analysis will allow us not only to place robust constraints on global sea level but also to identify the ``fingerprints'' produced by the gravitational, elastic, and isostatic effects of different melting ice sheets \citep[e.g., ][]{Mitrovica2001a}. It can thereby provide an independent test for different melt water sources in the Last Interglacial, and by extension the possible susceptibility of each ice sheet to future melting.

{
\begin{table*}[tbp]
\scriptsize
\centering
\caption{Sites, Number, and Types of Sea Level Indicators in the LIG Database} \label{table:sites}

\begin{tabular}{l l l r  } \scriptsize
\\ \hline
\textbf{Site} & \textbf{\# Obs.} & \textbf{Type} & \textbf{ Reference} \\
\hline
\multicolumn{3}{l}{\emph{Global}} \\
Global oxygen isotope stack & 51 & isotopic & \citet{Lisiecki2005a} \\[11pt]

\multicolumn{3}{l}{\emph{Northeastern Atlantic Ocean and Mediterranean Sea}} \\
Southern England & 2 & erosional & \citet{Westaway2006a} \\
Bristol Channel, Britain & 1 & erosional & \citet{Allen2002a} \\
Belle Hogue Cave, Jersey & 1 & erosinal & \citet{Keen1981a} \\
Port-Racine Beach, France & 1 & erosional & \citet{Bates2003a} \\
The Netherlands & 8 & facies & \citet{Zagwijn1983a} \\
Hergla South, Tunisia & 2 & facies & \citet{Hearty2007b} \\
Quaternary Basin, Mauretania & 2 & facies & \citet{Giresse2000a} \\[11pt]

\multicolumn{3}{l}{\emph{Northwestern Atlantic Ocean and Carribean Sea}} \\
Cape George, Nova Scotia & 1 & erosional & \citet{Stea1998a, Stea2001a} \\
Mark Clark, South Carolina & 1 & facies & \citet{Cronin1981a} \\
Grape Bay, Bermuda & 2 & facies & \citet{Muhs2002a, Hearty2007b} \\
San Salvador Island, Bahamas & 3 & reef & \citet{Chen1991a} \\
Great Inagua Island, Bahamas & 3 & reef; erosional & \citet{Chen1991a} \\
Abaco Island, Bahamas & 3 & reef; erosional & \citet{Hearty2007b} \\
Southern Barbados & 8 & reef & \citet{Schellmann2004a} \\[11pt]

\multicolumn{3}{l}{\emph{Southwestern Atlantic Ocean}} \\
Rio Grande do Sol coastal plain, Brazil & 1 & facies & \citet{Tomazelli2007a} \\
Camarones, Patagonia, Argentina & 1 & erosional &  \citet{Rostami2000a} \\[11pt]

\multicolumn{3}{l}{\emph{Pacific Ocean}} \\
Oahu, Hawaii & 3 & reef; corals; facies & \citet{Hearty2007b, Muhs2002a} \\
Mururoa Atoll & 1 & corals & \citet{Camoin2001a} \\[11pt]

\multicolumn{3}{l}{\emph{Australia}} \\
Eyre Peninsula & 1 & facies & \citet{Murray-Wallace1991a} \\
Rottnest Island & 1 & reef & \citet{Stirling1995a, Hearty2007b} \\
Minim Cove & 1 & facies & \citet{Hearty2007b} \\
Cape Range & 2 & reef & \citet{Stirling1998a} \\
Houtman Abrohlos Islands & 8 & reef; facies; corals & \citet{Zhu1993a, Eisenhauer1996a} \\[11pt]

\multicolumn{3}{l}{\emph{Indian Ocean and Red Sea}} \\
Red Sea & 30 & isotopic & \citet{Rohling2008a} \\
KwaZulu-Natal, South Africa & 3 & erosional; facies & \citet{Hobday1975a, Ramsay2002a} \\
Eastern Cape, South Africa & 1 & erosional & \citet{Ramsay2002a} \\
Maldives Archipelago & 1 & facies & \citet{Woodroffe2005a} \\
La Digue Island, Seychelles & 2 & reef & \citet{Israelson1999a} \\
Aldabra Atoll, Seychelles & 3 & corals; facies &\citet{Braithwaite1973a} \\[11pt]

\multicolumn{3}{l}{\emph{Polar regions}} \\
Northern and Western Alaska & 3 & facies & \citet{Brigham-Grette1995a} \\
Wrangel Island, Siberia & 1 & facies  & \citet{Gualtieri2003a} \\
Western Spitsbergen & 3 & erosional & \citet{Forman1984a,Andersson1999a} \\
Scoresby Sund, Greenland & 3 & facies & \citet{Landvik1994a, Vosgreau1994a} \\
Cape Ross, Antarctica & 1 & erosional & \citet{Gardner2006a} \\
\hline
\end{tabular}
\end{table*}
}

\section{Database of LIG Sea Level Indicators}

We characterize each sea level indicator in our database by five parameters: its geographical position, its altitude with respect to mean tide level, its age, the range of depths at which it might have formed, and the local uplift or subsidence rate. With the exception of geographical position, each of these variables has uncertainties that we assume follow a Gaussian distribution. For some values, including all depositional depth ranges, uniform distributions between two limits $a$ and $b$ may be a better choice than Gaussian ones. In these cases, we substitute a Gaussian distribution with the same mean and standard deviation as the uniform distribution, i.e. $(b-a)/\sqrt{12}$.

The full database is available on request from the authors and will be published with the final version of this paper.

\subsection{Nature of the indicators and depositional ranges}

The sea level indicators take a variety of forms, including: constructional coral terraces that provide both geomorphological and ecological information; coral biofacies in limestones that provide ecological but not geomorphological information; erosional features such as wave-cut terraces, sea caves, bioerosional notches, and raised beaches; and sedimentological and biofacial indicators of depositional depth.

Most of the indicators reflect deposition or formation within a specific range of depths. The most common reef terraces and associated coral assemblages, for instance, are generally interpreted as indicating deposition between mean low tide level and 5 m below mean low tide level \citep{Lighty1982a, Camoin2001a}. Intertidal sedimentary facies indicate deposition within the tidal range. While recognizing that LIG tidal amplitudes could have been slightly different than today, we convert descriptive ranges such as these into a common reference frame based on the tidal ranges reported in tide tables at a nearby modern locality. We also attempt to correct for variability in the measurement datum; while most sea level indicators have altitudes reported with respect to ``modern sea level'', some are more usefully described with reference to datums such as the mean low tide level or mean high tide level. We convert such datums into a mean tide level datum.

Some data, such as subtidal sedimentary facies, are limiting points; they place an upper or lower limit on past sea level but do not indicate a specific depositional depth. In statistical terminology, limiting points are censored data. 

\subsection{Age}

Age constraints on our data come from a variety of sources with a range of precisions. In some cases, age is constrained only by stratigraphic relationships with other units. In many cases, particularly involving coral reefs, radiometric (U/Th) dates are available. Other age constraints are derived from amino acid racemization, electron spin resonance dating, and related techniques such as thermoluminescence.

In three cases (the global oxygen isotope curve, the Red Sea oxygen isotope curve, and the Dutch sea level curve), relative ages are known with more precision than absolute ones.  As described in the Appendix, we have scaled and shifted the age models of the Red Sea and Dutch local sea level curves to be consistent with the \citet{Lisiecki2005a} age model for the global oxygen isotope curve. All of the dates outputted by our analysis should therefore be viewed within the context of this age model, which places the start of the Penultimate Termination at 135~ka and the peak of the Last Interglacial at about 122--126~ka.

When only a single conventional U/Th measurement from a unit is available, we expand the quoted ranges by 350\%, following the empirical observation of \citet{Scholz2007a} of the overestimate of the precision of ages from single sample measurements. When multiple measurements are reported, we employ their inverse-variance weighted mean. We expand the inverse-variance weighted standard deviation using a Student's $t$-distribution so that the 95\% confidence interval spans $\pm 1.96\sigma$, with $\sigma$ the standard deviation, as in a Gaussian distribution. 

\subsection{Tectonic uplift or thermal subsidence rate}

In order to remove the local tectonic contribution to paleo-sea level, we seek locally calibrated subsidence or uplift estimates for each locality. For most of the points in our database, no estimate of uplift or subsidence is available, but the value is expected to be near zero. For these location, we adopt an estimate for these locations of $0 \pm 1$ cm/ky. In a few regions where estimates are available, including much of the Bahamas and Hawai`i, subsidence or uplift is on the order of 1--2 cm/ky. A few localities have exhibited uplift (Barbados, Patagonia, southern England) or subsidence (the Netherlands, Pacific and Indian Ocean atolls) in excess of about 10 cm/ky.  The fastest uplifting locality in our database, Barbados, is rising at about 28 cm/ky.

\subsection{Coverage}

\begin{figure}[tbp] 
   \centering
   \includegraphics[width=3.25in]{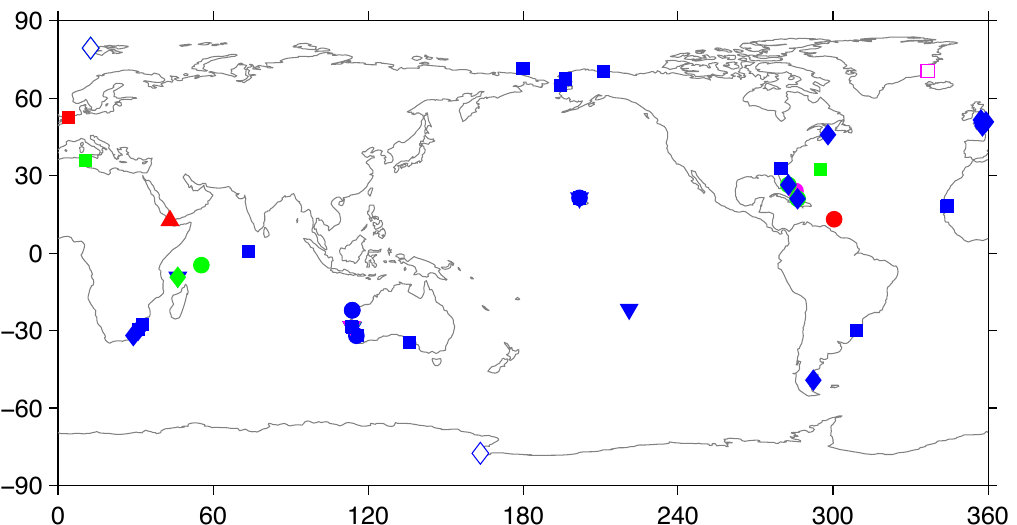} 
   \caption{Sites with at least one sea level indicator in our database. The symbol shapes reflect the nature of the indicators (upward triangles: isotopic; circles: reef terraces; downward triangles: coral biofacies; squares: sedimentary facies and non-coral biofacies; diamonds: erosional). The colors reflect the number of observations at a site (blue: 1; green: 2; magenta: 3; red: 4 or more). Locations marked by open symbols were excluded from our analysis because they have experienced strong near-field isostatic uplift for which our model cannot account.}
   \label{fig:sitemap}
\end{figure}

\begin{figure*}[bt] 
   \centering
   \includegraphics[width=6.5in]{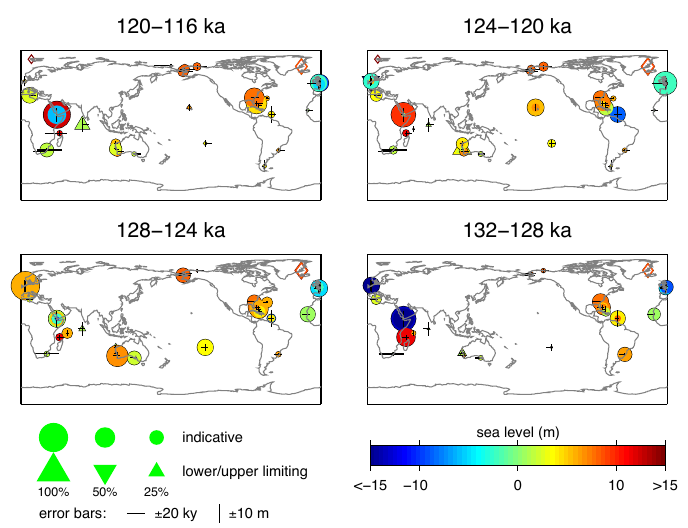} 
   \caption{Localities at which local sea level data exist in our database for time slices through the Last Interglacial. Points scale proportionately to the probability that they occur in the indicated interval. The horizontal lines are proportional to the standard deviation of the age measurement, and the vertical lines are proportional to the standard deviation of the sea level measurement . Censored data are indicated by upward (marine limiting) and downward (freshwater limiting) triangles. Color indicates the mean sea level estimate in meters. Open diamonds show near-field data points excluded from the analysis.}
   \label{fig:coveragemap}
\end{figure*}

Our database attains fairly good geographic coverage, including the northwestern, northeastern, and southwestern Atlantic coasts; the Carribean; Alaska, Greenland, Svalbard, and Siberia; Australia; the southwestern Indian coast; and Pacific and Indian Ocean islands (Figures \ref{fig:sitemap} and \ref{fig:coveragemap}; Table \ref{table:sites}). (Because the physical model we employ for developing our covariance function does not account for the local effects of isostatic rebound, we were, however, unable to include near-field data from Greenland, Svalbard, and Antarctica in our analysis.) Where nearby localities subject to less uplift are available, we have tried to limit the amount of data from rapidly uplifting sites, though we include Barbados because of its prominence in the literature. However, given the long history of the geological study of Pleistocene sea level indicators , which began not long after the collapse of the Diluvian hypothesis in the early nineteenth century \citep[e.g., ][]{Godwin-Austen1856a}, we do not claim that our database comprehensively represents the entire literature.

\begin{figure}[tbp] 
   \centering
   \includegraphics[width=3.25in]{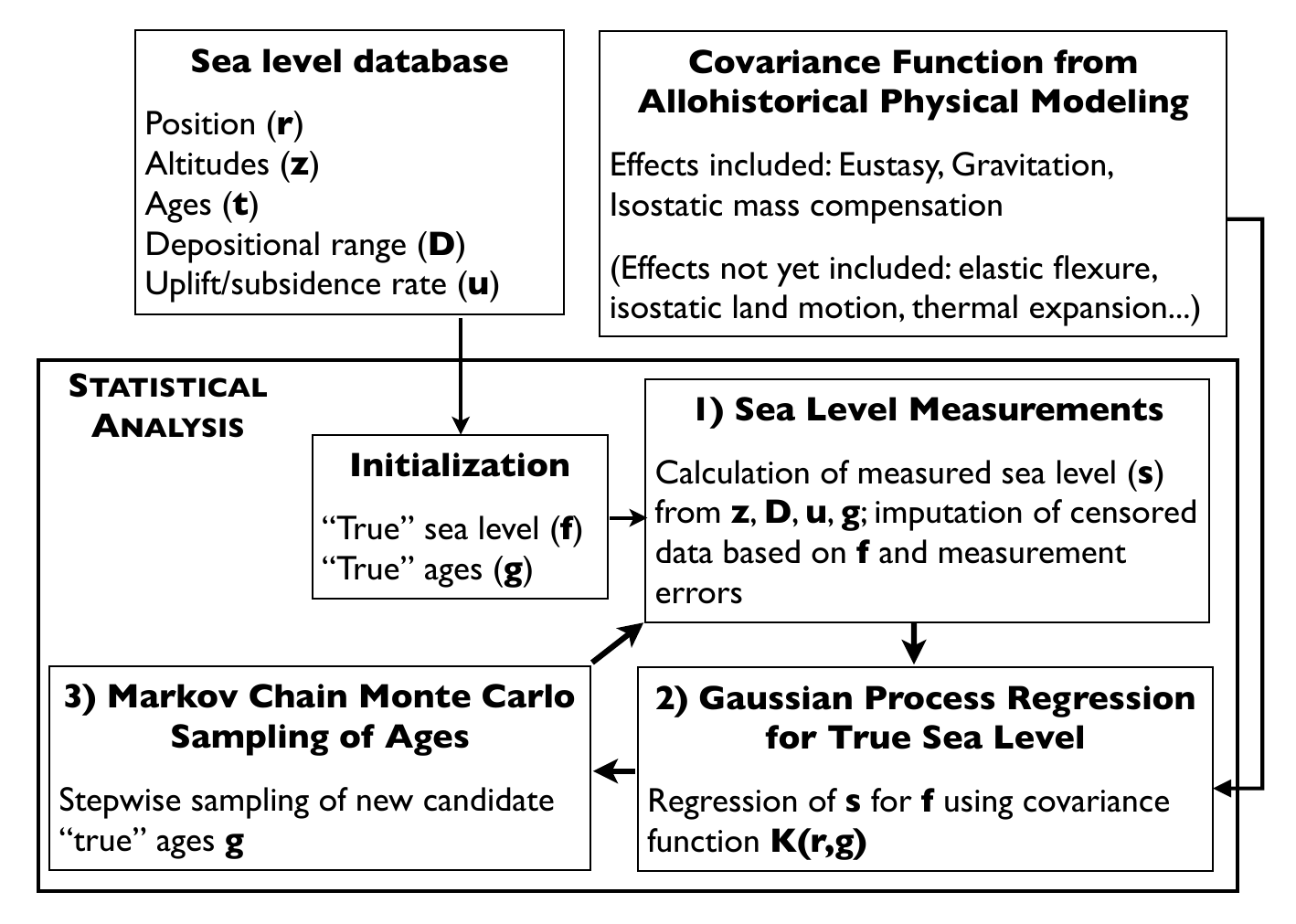} 
   \caption{Schematic illustration of the process used in our analysis.}
   \label{fig:processschematic}
\end{figure}

\section{Statistical Model}

\subsection{Preliminaries and Notation}

The ultimate goal of our statistical analysis is to determine the posterior probability distribution of sea level through time, conditioned upon the measurements in our database. Expressed symbolically, our aim is to evaluate the probability $P( f(\mathbf{x}, g) | \mathbf{r}, \mathbf{z}, \mathbf{t},\mathbf{D}, \mathbf{u})$ for locations $\mathbf{x}$ on Earth's surface and times $g$, where $f$ represents the true value of sea level at $\mathbf{x}$ and $g$. In our database, each sea level indicator is assigned an index $i = 1,\dotsc, N$ and is characterized by
\begin{description}
\setlength{\itemindent}{0pt}%
\item[$\mathbf{r}_i,$]{its exact geographic position,}
\item[$z_i,$]{a noisy measurement of its altitude,}
\item[$t_i,$]{a noisy measurement of its age,}
\item[$\mathbf{D}_i,$]{a closed or open interval reflecting its depositional range, and}
\item[$u_i,$]{a noisy estimate of the long-term average uplift or subsidence rate.}
\end{description}
When $\mathbf{D}_i$ is a closed interval, we replace it with $d_i$, a Gaussian estimate of depositional depth characterized by the same mean and variance as the uniform distribution on $\mathbf{D}_i$, as discussed before.

We collect these parameters into vectors $\mathbf{r}$, $\mathbf{z}$, $\mathbf{t}$, $\mathbf{D}$, $\mathbf{u}$, and $\mathbf{d}$. Similarly, we collect what will be the true sea levels in a vector $\mathbf{f}$ evaluated at the times $\mathbf{g}$ and locations $\mathbf{x}$, whose elements $f_j$, $g_j$ and $\mathbf{x}_j$ for $j = 1, \dotsc, M$ are the desired sea levels and evaluation points. Only when geographical positions and depositional ranges are concerned does the bold vector notation serve double-duty: $\mathbf{x}$ and $\mathbf{r}$ are either coordinates or vectors of coordinates, and $\mathbf{x}_i$, $\mathbf{r}_i$ and $\mathbf{x}_j$, $\mathbf{r}_j$ are individual sets of coordinates. Likewise, $\mathbf{D}$ is either a depositional range or an array of depositional ranges, and $\mathbf{D}_i$ is an individual depositional range. This dual purpose is not, however, likely to lead to confusion.

\subsection{Gaussian process regression}

We proceed from this point using a Gaussian process approach \citep{Rasmussen2006a}. We must select some covariance function for true sea level, $k(\mathbf{r}_i, g_i; \mathbf{r}_j, g_j)$, as we will address in section \ref{subsection:covariance}. Let $(\mathbf{f}, \mathbf{g})$ refer to the vectors of true sea levels and ages that correspond to the vectors of measurements $(\mathbf{z}, \mathbf{t}, \mathbf{D}, \mathbf{u})$; i.e., with every entry $(f_i, g_i)$, we associate an entry $(z_i, t_i, \mathbf{D}_i, u_i)$ for all indices $i=1, \dotsc, N$. With the covariance function $k$ given, we can then readily recover an estimate of true sea level  at any arbitrary location $\mathbf{x'}$ and time $g'$ through straight-forward kriging interpolation \citep{Press2007a}. We denote the mean and variance of this estimate by $\overline{f}(\mathbf{x'}, g')$ and $\mathbf{V}(f(\mathbf{x'}, g'))$, respectively.

As before, the vectors $\overline{\mathbf{f}}$, $\mathbf{x}'$ and $\mathbf{g}'$ will collect the mean estimates of the sea levels at the desired points $\mathbf{x}'$ and $\mathbf{g}'$ in space and time. The sets of desired evaluation points $(\mathbf{x}'_j, g'_j)$, $j=1, ..., M$, and the measurements $(\mathbf{r}_i, g_i)$, $i=1,...,N$ need not necessarily overlap. The matrix $\mathbf{V}''$ collects the kriging (co)variance of $\mathbf{f}'$ at and between $(\mathbf{x}', \mathbf{g}')$. Let $\mathbf{K}$, $\mathbf{K'}$, and $\mathbf{K''}$ be the covariances of $(\mathbf{f}, \mathbf{g})$ and/or $(\mathbf{f'}, \mathbf{g'})$ at the observed and desired points, i.e., let the symmetric square matrices $\mathbf{K}$ and $\mathbf{K''}$ and the rectangular matrix $\mathbf{K'}$ be defined by their elements:

\begin{align}
\label{krig1}
K_{ij} & = k(\mathbf{r}_i, g_i; \mathbf{r}_j, g_j) & \textrm{ where } i, j = 1, ..., N, \\
K''_{ij} & = k(\mathbf{x}'_i, g'_i; \mathbf{x}'_j, g'_j) &  \textrm{ where } i, j = 1, ..., M, \\
K'_{ij} & = k(\mathbf{r}_i, g_i; \mathbf{x}'_j, g'_j)  & \textrm{ where } i = 1, ..., N
\\ \notag & & \textrm{ and } j = 1, ..., M.
\end{align}
From this, the kriging step consists of calculating $\overline{\mathbf{f}}$, the $M\times 1$ vector of mean sea level estimates at $(\mathbf{x}', \mathbf{g}')$, as

\begin{align}
\label{krig2} \overline{\mathbf{f}} & =  \mathbf{K}'^\top \mathbf{K}^{-1} \mathbf{f},
\end{align}
which has

\begin{align}
\label{krig3} \mathbf{V''} & = \mathbf{K}'' - \mathbf{K}'^\top \mathbf{K}^{-1} \mathbf{K}'
\end{align}
as its $M \times M$ covariance matrix. It is clear from the above that, when $\mathbf{x}' = \mathbf{r}$ and $\mathbf{g'} = \mathbf{g}$, $\mathbf{K} = \mathbf{K}' = \mathbf{K}''$, and therefore $\overline{\mathbf{f}} = \mathbf{f}$ and $\mathbf{V''} = \mathbf{0}$. In other words, when the queried points are identical to the measurement locations, the interpolated values of true sea level remain unchanged and receive no kriging variance.

We can therefore replace the problem of finding the posterior probability of sea level anywhere, $P( f(\mathbf{x}, g) | \mathbf{r}, \mathbf{z}, \mathbf{t},\mathbf{D}, \mathbf{u})$, with the more tractable problem of finding $P(\mathbf{f}, \mathbf{g} | \mathbf{z}, \mathbf{t},\mathbf{D}, \mathbf{u})$, which is the posterior probability of sea level at the smaller set of points defined by the measurement locations. After adjusting altitude $z_i$ for uplift or subsidence rate $u_i$ over a time $g_i$, we define the corrected altitude $z'_i$  as

\begin{equation}\label{upliftcorrectedaltitutde}
z'_i \equiv z_i - g_iu_i,
\end{equation}
with variance

\begin{equation}
\sigma^2_{z'i} \equiv \sigma^2_{zi} + g_i^2\sigma^2_{ui},
\end{equation}
and we define the sea level measurement $s_i$ and its variance $\sigma^2_{si}$ as

\begin{align}\label{sea levelmeasurement}
s_i \equiv z'_i - d_i, \\
\sigma^2_{si} \equiv \sigma^2_{z'i} + \sigma^2_{di},
\end{align}
where $\sigma^2_{zi}$, $\sigma^2_{ui}$, and $\sigma^2_{di}$ are the variances respectively of altitude $z_i$, uplift rate $u_i$, and depositional depth $d_i$. By Bayes' theorem,

\begin{equation}\label{bayes}P( \mathbf{f}, \mathbf{g} | \mathbf{s}, \mathbf{t}) \propto P( \mathbf{s},  \mathbf{t} | \mathbf{f}, \mathbf{g}) \cdot P(\mathbf{f}, \mathbf{g}).\end{equation}
We drop the position variable $\mathbf{r}$ from the notation, since its values are fixed in the data set and implicit in the indexing of the other variables. For uncensored sea level measurements, we have the likelihood

\begin{equation}P(s_i | f_i, g_i) \sim \mathcal{N}(f_i, \sigma^2_{si}).\end{equation}
In other words, the probability of observing sea level $s_i$ at a point in the data set that has a true sea level of $f_i$ is given by a Gaussian centered on the truth with variance $\sigma^2_{si}$.
For censored data,

\begin{equation}\label{censoredsdist}P(s_i | f_i, g_i) \sim \mathcal{N}(f_i, \sigma^2_{si}) \cdot \delta\big((z'_i - s_i) \in \mathbf{D}_i\big) \end{equation}
where  $\delta$ is an indicator function that is 1 when $z'_i - s_i$ is in the depositional range $\mathbf{D_i}$ and 0 otherwise. For instance, if $\mathbf{D}_i$ is $(-\infty, -2]$, reflecting deposition at least two meters below mean tide level, then $\delta$ would be 1 for $s_i \geqslant z'_i + 2$ and 0 otherwise. For age measurements, we have the likelihood

\begin{equation}\label{tdist}P(t_i | g_i)  \sim \mathcal{N}(g_i, \sigma^2_{ti}),\end{equation}
where $\sigma_{ti}^2$ is the variance of age measurement $t_i$.
For the sea level vector $\mathbf{f}$, we assume the prior

\begin{equation}\label{fprior}P(\mathbf{f}|\mathbf{g})  \sim \mathcal{N}(\mathbf{0}, \mathbf{K}(\mathbf{g})),
\end{equation}
where we use the notation $\mathbf{K}(\mathbf{g})$ for the covariance to emphasize its dependence on ages $\mathbf{g}$. For the age vector $\mathbf{g}$ itself, we assume a uniform prior.

\subsection{Algorithm for sampling the sea level distribution}

To explore the distribution in equation \ref{bayes}, we use a recursive three-step algorithm (schematically illustrated in Figure \ref{fig:processschematic}) to generate updates of $\mathbf{f}$, $\mathbf{g}$, and $\mathbf{s}$. We start by initializing $\mathbf{g} = \mathbf{t}$ for all data points and $z'_i = z_i - g_iu_i$ and $f_i = s_i = z'_i - d_i$ for the uncensored ones. By simple kriging interpolation (equations \ref{krig2} and \ref{krig3}), we estimate $f_i$ at the remaining data points.

1. In step one of our algorithm, we calculate values of sea level measurements $\mathbf{s}$ from $\mathbf{z}$, $\mathbf{D}$, $\mathbf{g}$ and $\mathbf{u}$. For uncensored data, $s_i$ is as defined in equation \ref{sea levelmeasurement}. For censored data, we sample $s_i$ from the distribution in equation \ref{censoredsdist}, with an additional variance term $\sigma_{fi}^2$, the kriging variance of $f_i$.

2. In step two, we update our estimate of true sea level $\mathbf{f}$ based upon the new $\mathbf{s}$ as follows. We define the matrix of the sea level measurement noise $\mathbf{N}$, with elements $\sigma^2_{si}$ along the diagonal and zero elsewhere. Then, by Gaussian process regression, paralleling equation \ref{krig2},  we calculate 

\begin{equation}
\mathbf{f}  = \mathbf{K}(\mathbf{g})^{\top} (\mathbf{K(g) + N})^{-1} \mathbf{s}, \end{equation}
the vector of sea level predictions and the vector of their variances

\begin{equation}
\mathbf{\Sigma}  = \textrm{diag} \{ \mathbf{K(g)}^{\top} (\mathbf{I} -  (\mathbf{K(g) + N})^{-1} \mathbf{K}(\mathbf{g}) ) \},
\end{equation}
where $\textrm{diag}$ denotes the diagonal elements.  

3. In step three, we update our estimate of the true ages $\mathbf{g}$. To do this, we follow a Markov Chain Monte Carlo approach applying the Metropolis-Hastings algorithm sequentially to each $g_i$. Let $\mathbf{g}_{-i}$ represent $\mathbf{g}$ with element $i$ removed. For each $i$, we sample from the distribution $P(g_i | \mathbf{t}, \mathbf{g}_{-i}, \mathbf{f} )$, which, by multiple applications of Bayes' theorem and the facts that $P(\mathbf{t} | \mathbf{g}) =  \prod_{i}  P(t_i | g_i)$ and that $P(\mathbf{t} | \mathbf{f}) = P(\mathbf{t})$, reduces as

\begin{equation}
\begin{split}
P(g_i | \mathbf{t}, \mathbf{g}_{-i}, \mathbf{f} ) 
& \propto P(t_i | g_i) \cdot P(\mathbf{f} | \mathbf{g}) \cdot P(\mathbf{g}).
\end{split}
\end{equation}
The first term is given by equation \ref{tdist}, and the second term by equation \ref{fprior}.
We can drop the third term because of our assumption of a uniform prior for $\mathbf{g}$. 

We generate test values $g'_i$ using a Gaussian function $q(g'_i ; g_i)$ centered at $g_i$ and bounded such that, when stratigraphic ordering is known, a point $j$ that follows a point $i$ always has $g_j \leqslant g_i$. (Where no bounds apply, $q(a ; b) = q(b ; a)$.) For the sequences where relative ages are known more precisely than absolute ones, these are calculated in terms of time after the preceding point. Following the Metropolis-Hastings algorithm \citep{Hastings1970a}, we accept a candidate $g'_i$ with probability 

\begin{multline}
\label{eqn:acceptanceprob}
\min \left(1, \frac{P(g'_i | \mathbf{t}, \mathbf{g}_{-i}, \mathbf{f}) \cdot q(g_i ; g'_i)}{P(g_i | \mathbf{t}, \mathbf{g}_{-i}, \mathbf{f}) \cdot q(g'_i ; g_i)} \right) = \\ \min \left(1, \frac{P(t_i|g'_i) \cdot P(\mathbf{f}|\mathbf{g}_{-i},g'_i) \cdot q(g_i ; g'_i)}{P(t_i|g_i) \cdot P(\mathbf{f}|\mathbf{g}_{-i}, g_i) \cdot q(g'_i ; g_i)} \right).
\end{multline}
So that we can assess results within a common temporal reference frame, we arbitrarily set the temporal variance $\sigma_{ti}^2$ for the first step of our longest quasi-continuous sequence of data points (the sea level curve derived from the global oxygen isotope stack, for most runs) to zero. 

This algorithm, repeated a large number of times, samples the probability distribution described by equation \ref{bayes}. We thin the results by storing every 20th sample and account for burn-in by discarding the first 50 stored samples. After several parallel executions of the algorithm, each of which store at least about 200 samples, we check for convergence by inspecting the autocorrelation of stored values of $\mathbf{g}$ and discard executions that appear not to converge.
 To generate our target distribution $P( f(\mathbf{x}, g) | \mathbf{s}, \mathbf{r}, \mathbf{t})$, we use kriging interpolation (equations \ref{krig1}--\ref{krig3}) to estimate the sea level field at all spatial and temporal points of interest for each stored sample.
 
 We note that this algorithm, while satisfying from a theoretical perspective, could benefit from greater computational efficiency. The most time-consuming steps in its execution are the inversions of the covariance matrices, which for a database of $n$ samples require $\mathcal{O}(n^3)$ operations. This inversion occurs once in step 2 and $n + 1$ times in step 3. Thus, each iteration of the algorithm is $\mathcal{O}(n^4)$. Repeating the algorithm a few thousand times in the courses of a Monte Carlo simulation with a database of about 150 points can therefore take several days; without increased efficiency, larger data sets will become unmanageable. 

\subsection{The Covariance Function}
\label{subsection:covariance}

We use a covariance function that takes the form
\begin{equation}
k(\mathbf{r}_i, g_i; \mathbf{r}_j, g_j) = k_s(\mathbf{r}_i; \mathbf{r}_j) \cdot k_t(g_i; g_j) 
\end{equation}
where $k_s$ and $k_t$ are respectively the spatial and temporal components of the covariance function. To find suitable $k_s$ and $k_t$, we employ a simple physical model of the gravitational effects of ice sheet melting and isostatic mass compensation. To generate $k_s$, we run the model 300 times. During each run, we produce complete world maps of sea level at 20 random time slices, for a total of six thousand sea level maps. From these six thousand sea level maps, we compute the covariance among local sea levels at evenly spaced points on a Cartesian grid (spaced at 5$^\circ$ latitude and 10$^\circ$ longitude) and between these points, mean global sea level, and the volumes of the different ice sheets. We store the results as a lookup table, which is effectively over-sampled near the poles because of the use of the Cartesian grid. For the covariance between two arbitrary points, we reference the closest grid points.

Our physical model is based upon \citet{Woodward1888a} and the discussion thereof in \citet{Farrell1976a}. Given a change in ice volume at a point $p$ corresponding to global sea level rise of $m_i$,  the change in sea level $R_i$ at an angular distance of $\theta$ is given by

\begin{align}
\label{eqn:trainingmodel}
R_i(\theta) = m_i \cdot (1 + \ell(\theta)), \\
\ell(\theta) = \left\{1 - \frac{1}{2 \sin(\theta/2)} \right\} \left/ \frac{\rho_E}{3 \rho_w }, \right.
\end{align}
where $\rho_E$ is the mean density of the Earth (5.5 g/cm$^3$) and $\rho_w$ is that of seawater (1.0 g/cm$^3$). To prevent singularities, we do not let $R$ take values smaller than -10 m, a constraint equivalent to assuming that all points are at least 3$^\circ$ from a point mass. To approximate the gravitational effects of isostatic compensation, we place a slowly adjusting compensatory mass at $p$. The mass equivalent $m_c$, in units of length, is given by the differential equation

\begin{equation}
\frac{d(m_c - m_i)}{dt} = \frac{m_i - m_c}{\tau_c},
\end{equation}
where $\tau_c$ is the timescale of isostatic adjustment. This approach does not take into account the effects of flexure or isostasy on the solid Earth and so captures near-field behavior only crudely. In each model run, $\tau_c$ is drawn from a uniform distribution between 0.1 and 14.1 ky.  We include the low end of the distribution not because they are physically plausible isostatic timescales but because they allow us to include in the prior probability distribution scenarios in which eustasy completely dominates other processes.

\begin{figure*}[tb] 
   \centering
   \includegraphics[width=6.5in]{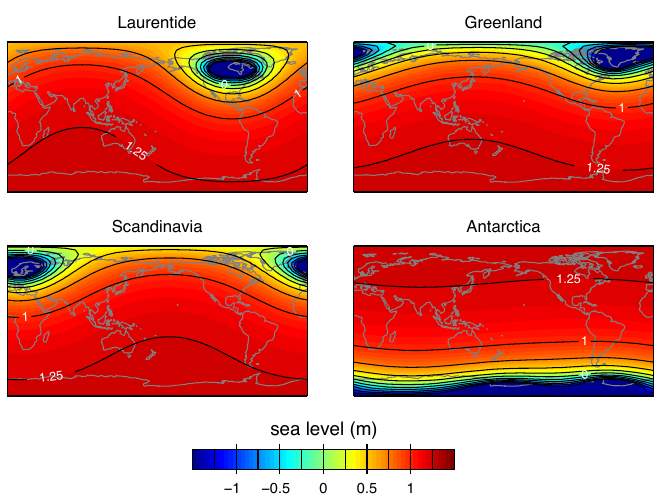} 
   \caption{Gravitational fingerprints of the (a) Laurentide, (b) Greenland, (c) Scandinavian, and (d) Antarctic ice sheets used in generating the covariance function.}
   \label{fig:fingerprints}
\end{figure*}

Including the gravitational effects of this compensatory mass, the change in sea level $R$ at $\theta$ is given by 

\begin{equation}
R(\theta) = m_i + (m_i - m_c) \cdot \ell(\theta).
\end{equation}
We treat changes in sea level as resulting from changes in the mass of four ice sheets, representing the Laurentide (represented as two point ice sheets at 57.2$^\circ$ N, 102.2$^\circ$W and 56.5$^\circ$N, 78.5$^\circ$W), Greenland (at 65.5$^\circ$N, 49.5$^\circ$W and 76.2$^\circ$N, 22.7$^\circ$W), Scandinavian (at 64.2$^\circ$N, 14.5$^\circ$E), and Antarctic (at 81.5$^\circ$S, 176.5$^\circ$W and 77.5$^\circ$S, 52.5$^\circ$W) ice sheets. These masses give rise to the gravitational fingerprints shown in Figure \ref{fig:fingerprints}. With long isostatic compensation timescales, these fingerprints are strongly expressed; with short isostatic compensation time scales, local sea level is nearly equal to global sea level.

\begin{figure}[tb] 
   \centering
   \includegraphics[width=3.25in]{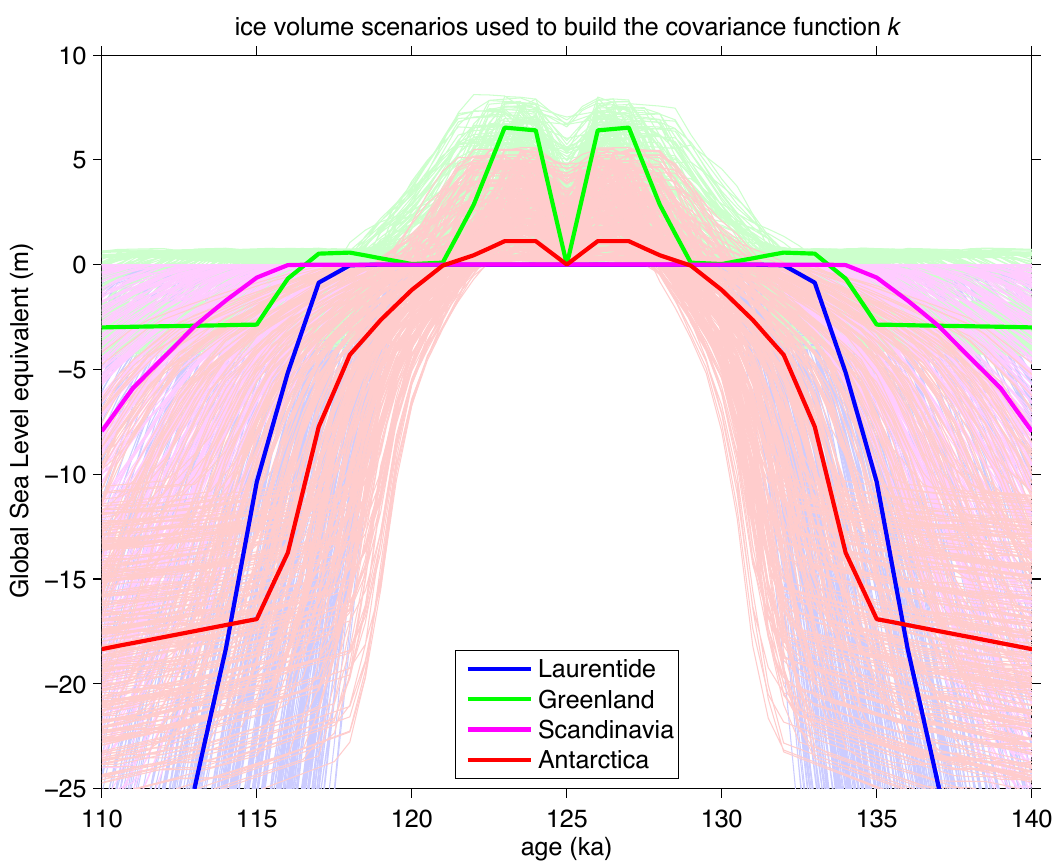} 
   \caption{Synthetic ice volume histories, expressed in terms of global sea level equivalent, used to build the covariance function. The histories are based on random distortions of the ICE-5G reconstruction of LGM-to-modern ice volume. The pale curves illustrate the distribution of alternative histories used to generate the covariance function. The bold curves are used to generate the synthetic data for the validation analysis.}
   \label{fig:icehistory}
\end{figure}

To generate reasonable ice sheet histories, we employ modifications of a modified form of the ICE-5G ice sheet history \citep{Peltier2004a} (Figure \ref{fig:icehistory}). In the base history, the ICE-5G ice sheet history from 21~ka to present is transposed to 146 to 125~ka, and occurs in reverse from 125 to 104~ka. To generate a new scenario, the amplitudes of melting associated with the Greenland, European, North American, and Antarctic ice sheets are each multiplied by uniformly distributed random numbers between 0.6 and 1.4.  For each ice sheet, the 21 ky timescale of melting is stretched by a uniformly distributed random factor drawn between 0.5 and 1.5 during the approach to the peak interglacial, and stretched by another random factor from the same distribution after the peak interglacial.

In addition, we allow for up to 5 m of additional melting from Antarctica between 127 and 123~ka, and up to 7 m of melting from Greenland between 127 and 123~ka. The magnitudes are drawn from uniform random distributions. As with the primary ice sheet melting, the time scales for the additional melting are stretched by random factors between 0.5 and 1.5. We also allow a brief lowstand at 125~ka caused by the loss of between 0\% and 100\% of the additional melting from each ice sheet.

\begin{figure}[tb] 
   \centering
   \includegraphics[width=3.25in]{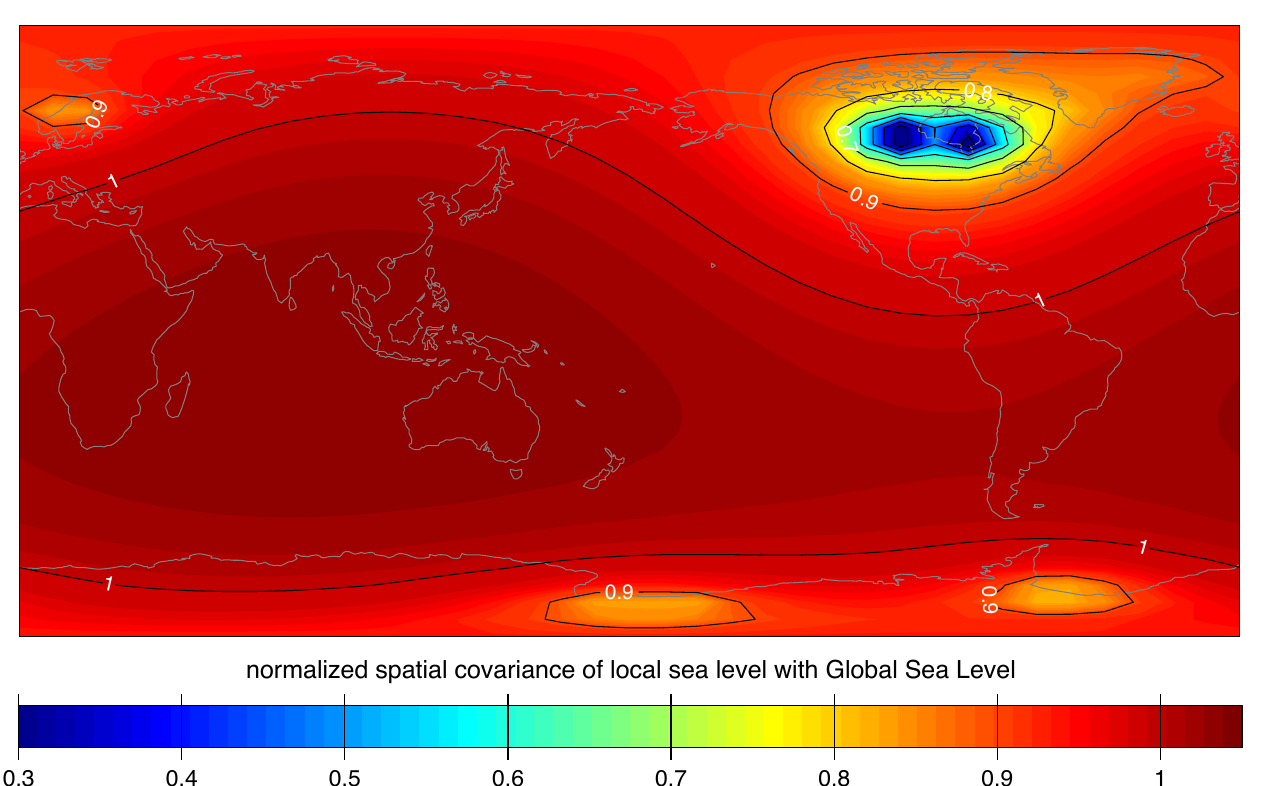} 
   \caption{The spatial covariance of local sea level with global sea level $k_s(GSL; \mathbf{r}_i)$, normalized to the variance of global sea level ($\sigma_{GSL}$ = 33.4 m).}
   \label{fig:spatialcov}
\end{figure}

While all sea levels are measured with respect to the modern sea level datum, changes in ice mass since the Last Glacial Maximum have not yet been fully compensated. In order for our model to report sea levels in this datum,  it is therefore necessary to calculate the difference between modern sea level and equilibrium sea level. For a given isostatic timescale $\tau_c$, we assume that ice masses were fully compensated (i.e., the mantle and ice sheets were in equilibrium) at 21~ka  and follow the ICE-5G ice sheet histories to calculate modern sea level. In our random model runs, we assume that ice sheets are fully compensated at peak glaciation and subtract modern sea levels from all values. One portion of the resulting spatial covariance function, the covariance of local sea level with global sea level, is shown in Figure \ref{fig:spatialcov}.

To find a suitable temporal covariance function $k_t$, we explicitly calculate the temporal covariance function for global sea level and fit it using the sum of two Gaussian curves:

\begin{equation}
k_t(g_i; g_j) = \sum_{n=1}^2 f_n \exp\left[-\left(\frac{g_i - g_j}{\tau_{k,n}}\right)^2\right]
\end{equation}
The best fitting values are $\tau_{k,1} = 4.31$ ky, $\tau_{k,2} = 0.41$ ky, $f_1 = 0.97$, and $f_2$ = 0.03 (Figure \ref{fig:temporalcov}).

Although our physical model is extremely simplified, neglecting all changes in the shape of the solid Earth as well as dynamic and rotational effects, and although coupling our statistical approach to a more sophisticated model may be useful, we do not think these simplifications will significantly alter our global sea level projections. We use our physical model to construct a reasonable prior probability distribution that discriminates regions where local sea level is strongly correlated with global sea level from regions where it is less strongly correlated with global sea level. The simple model serves this purpose. Its failings may be more acute in evaluating local sea level in the near-field of ice sheets, where the neglected changes to the shape of the solid Earth can be the dominant factor in sea level change.

\subsection{Validation of method using synthetic data}

To test our statistical model, we generated a synthetic sea level history in the same fashion as during the covariance calibration process, using our base model with the arbitrary addition of 6.6 additional meters of melt from Greenland and 1.1 meters additional melt from Antarctica. We sampled the history at the same points in space and time (case A) as sampled by our database, under two conditions: (1) with no errors in ages and 10 cm errors in sea level; (2) with the same chronological and sea level errors as in the data set. We also ran conditions (1) and (2) excluding the oxygen-isotope derived global sea level curve (case B) and excluding both the oxygen-isotope curve and the Red Sea sea level curve of \citet{Rohling2008a} (case C). We discuss the results of the validation analysis in section \ref{sect:results_validation}.

\subsection{Summary Statistics}
\label{sect:summarystats}

We report several summary statistics for each four-dimensional sea level distribution we discuss. We assess the \emph{median} and \emph{quantiles of the data points} by aggregating the sub-distributions determined by the stored mean and standard deviations of sea level data points and ages for each stored iteration of the MCMC model.

We similarly assess the \emph{median} and \emph{quantiles of global sea level} by aggregating sub-distributions for global sea level over time from each stored iteration of the MCMC model. Each sub-distribution is determined from the stored mean and standard deviations of sea level data points and ages associated with the iteration by Gaussian process approximation of GSL at 500-year intervals from 115 to 140~ka.

We calculate the \emph{1000-year and 500-year average rates of global sea level change} by taking the average slope of the global sea level curve from each iteration over 1000-year intervals or 500-year intervals and aggregating these curves to produce a distribution of global sea level rate.  Note that these rates are average rates over several centuries; they place a lower bound on century-level rates of sea level rise.

Of particular interest are the highest global sea level reached during the Last Interglacial and the fastest rate of sea level rise experienced. We report two sets of statistics relevant to these questions. First, we report the \emph{maximum of the median global sea level curve and sea level rate curve} and its confidence interval. We also compute \emph{global sea level and global sea level rate exceedance probabilities.} To do this, we sample each sub-distribution of global sea level one hundred times and aggregate all of these samples. In order to discount time points at which we have limited data, we incorporate only the time points within each sample at which the posterior standard deviation is less than 15\% of the prior standard deviation. We then identify the fraction of sample histories in which a 1000-year running average of GSL or the 1000-year or 500-year average rate exceeds a given value. The 95\% probability exceedance levels, for instance, are the values that with 95\% confidence we can say the sea level or sea level rate exceeded. For the rate maxima and rate exceedance probabilities, we focus on intervals beginning when global sea level was -10 m or higher, as we expect that ice sheet dynamics during these intervals will more closely resemble future ice sheet dynamics than will the behavior of ice sheets during intervals of lower GSL.

We compute parallel summary statistics for Northern Hemisphere and Southern Hemisphere ice volume, arrived at by Gaussian process regression for these values instead of for GSL.

To identify outliers among the data points, we compute the \emph{probability of a measurement} given the assessed sea level distribution. To do this, we take the average over all $N$ stored MCMC iterations of the probability that the parameter $f$ (local sea level, global sea level, or age) with measured value $f_m \pm \sigma_m$ was drawn from the distribution indicated by iteration $i$, with mean $f_i$ and standard  deviation $\sigma_i$.  For indicative points, the probability for each iteration is given by a $\chi^2$ distribution with one degree of freedom on the parameter $\frac{(f_i - f_m)^2}{\sigma_i^2 + \sigma_m^2}$. For limiting points, the probability is given by a cumulative normal distribution with mean $f_i - f_m$ and variance $\sigma_i^2 + \sigma_m^2$.

To compare different probability distributions $\mathbf{f}_1$ and $\mathbf{f}_2$ for a parameter $f$ computed using different subsets of the data, we calculate the \emph{expected Mahalanobis distance}. We sample with replacement 1000 pairs of MCMC iterations $(\mathbf{f}_{1,i},\mathbf{f}_{2,i})$ from $\mathbf{f}_1$ and $\mathbf{f}_2$. We then take the mean Mahalanobis distance between each pair. The Mahalanobis distance of each pair is given by $\sqrt{(\mathbf{f}_{1,i} - \mathbf{f}_{2,i})^\top (\mathbf{\Sigma}_{1,i} + \mathbf{\Sigma}_{2,i}) (\mathbf{f}_{1,i} - \mathbf{f}_{2,i})}$ where $\mathbf{\Sigma}_{j,i}$ is the variance of $\mathbf{f}_{j,i}$.

\section{Results}

\subsection{Validation analysis  \label{sect:results_validation}}

\begin{figure*}[tb] 
   \centering
   \includegraphics[width=6.5in]{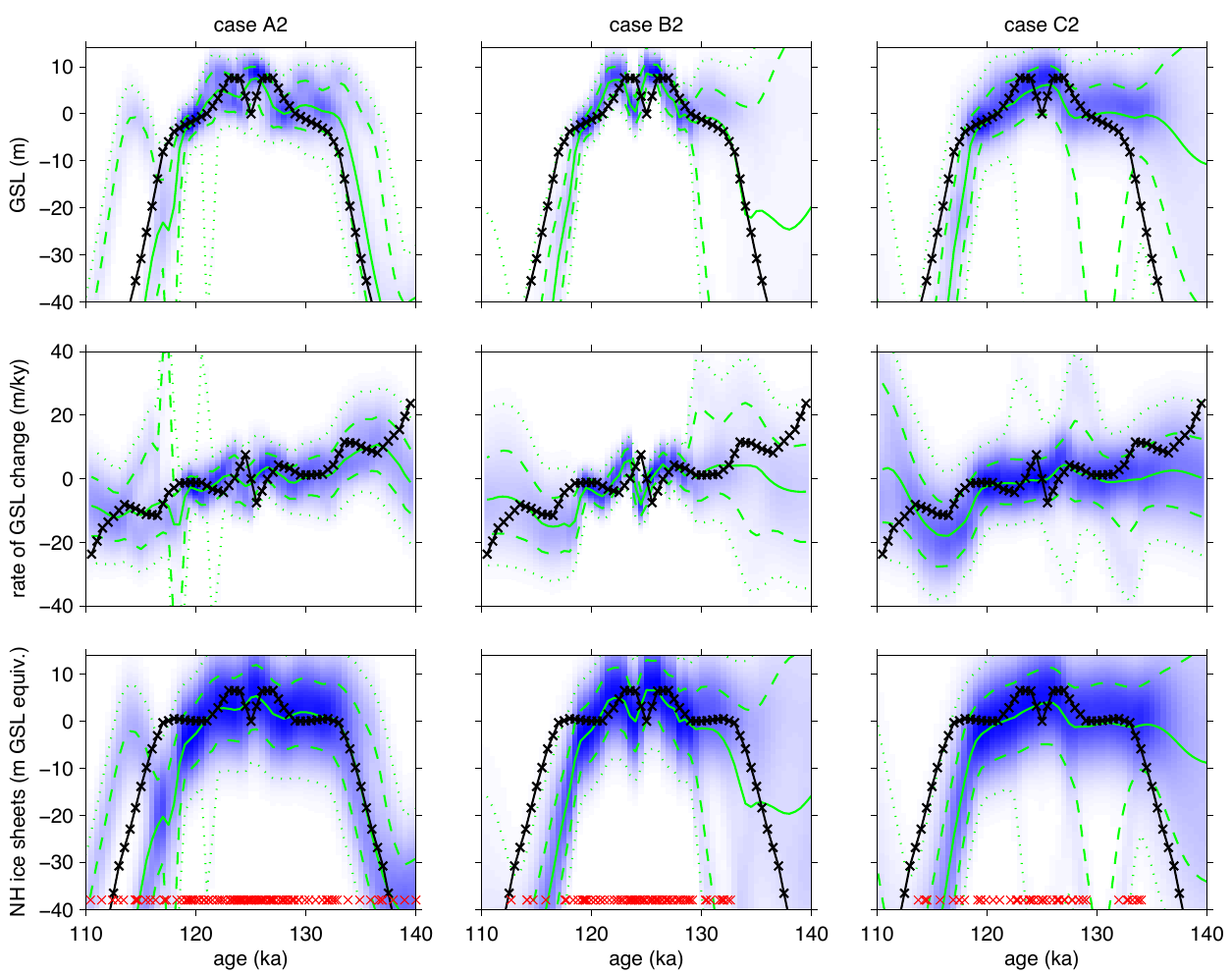} 
   \caption{Probability density plots (blue) of global sea level (top row), 1000-year average global sea level rates (middle row) and Northern Hemisphere ice volume (bottom row) as recovered by sampling a synthetic data set (black).  The synthetic data was sampled with the same errors as in the real data set. Case A2 includes all data, case B2 excludes the global $\delta^{18}O$ curve, and case C2 excludes both the global $\delta^{18}O$ and the Red Sea data. The solid line marks the median projection, dashed lines mark the 16th and 84th percentiles, and dotted lines mark the 2.5th and 97.5th percentiles. Crosses at bottom mark the median posterior estimates of the sample ages.}
   \label{fig:validation}
\end{figure*}

Comparisons of true and projected global sea level, rate of global sea level change, and Northern Hemisphere ice volume for the validation analysis are shown in Figure \ref{fig:validation}. (The cases with no sampling errors are shown in figure \ref{fig:validation_noerrors}). Summary statistics are presented in Tables \ref{table:validationstats_gsl}. We show both the 95\% probability exceedance levels and the exceedance probabilities corresponding to the true maximum values.  The algorithm performs a good job of reconstructing global sea level, with the median projection often quite close to the true values, generally within the 67\% confidence intervals, and always within the 95\% confidence intervals. The same is true for rates and ice volumes.

Cases A1, B1, and C1  highlight the interpretive limitations imposed by sampling (Tables \ref{table:validationstats_gsl} and Figure \ref{fig:validation_noerrors}). All of these ``perfect knowledge'' cases do a good job of reconstructing global sea level and rates for the period of highest sea level. The maximum of the median projections for GSL and rate are quite close to the true maxima, which fall between the 58\% and 72\% probability exceedance levels. The true rates of change fall between the 23\% and 64\% probability exceedance values. Without the $\delta^{18}$O curve included (as in cases B1 and C1), the resolution of the curves becomes poor before about 130~ka, but in all cases the reconstructions do a good job of resolving details, including the 125~ka drawdown, after 130~ka. However, even under the best of circumstances, we do a mediocre job of reconstructing ice volumes; the 67\% confidence intervals for our maximum ice volume projections span about $\pm 6$ m. The true NH ice volumes fall between the 69\% ans 72\% probability exceedance levels, while the true SH ice volume falls between the 92\% and 94\% probability levels.

In cases A2, B2, and C2, where the measurement errors are included, they prevent resolution of some of the details of the curves, including the brief 125~ka drawdown (Tables \ref{table:validationstats_gsl} and Figure \ref{fig:validation}). Case B2 preserves a suggestion of the drawdown, while case A2 resolves resolves multiple wobbles of sea level instead of a single drawdown. Case C2 finds a smoother sea level peak. Compared to cases A1--C1, these cases exhibit increased uncertainty in estimates of the maximum of the median GSL and GSL rate curve, but preserve accuracy, with the true values remaining well within the 67\% confidence interval. Compared to the more accurate cases, the exceedance levels are biased toward higher values. The true value of the GSL maximum falls between the 79\% and 91\% probability exceedance levels, while the true value of the rate of change falls between the 83\% and 91\% probability exceedance values. The precision and accuracy of the ice volume projections are comparable to those of cases A1--C1, indicating that insufficient sampling rather than measurement error is the major source of uncertainty in the ice projections. Based on the exceedance levels for these cases, we employ the 80\% and 95\% probability levels to bracket our estimates of maximum values in our analysis of the real data.

\begin{table*}[tbp]
\textbf{\caption{Statistics from Validation Analysis}}
\scriptsize
\renewcommand\arraystretch{1.3}
\begin{tabular}{|l|rr|r|r|rr|r|r|}
\hline
& \multicolumn{4}{c|}{Max. Global Sea Level} & \multicolumn{4}{c|}{GSL Rate at $\geq$ -10 m}\\ \hline

& \multicolumn{2}{c|}{Median Projection} & 95\% Exceed. & Prob. of &\multicolumn{2}{c|}{Median Projection} & 95\% Exceed.  & Prob. of \\
Case & Time (ka) & Max. (m) & Level (m) &  True Value & Time (ka) & Max. (m/ky) & Level (m/ky)  & True Value \\
\hline

True & 127 or 123 & 7.7  &   & & 124.5 & 7.5 & & \\ \hline
A1 & 123.5 & $7.6 \pm 0.6$ & 7.3 & 58\% & 124.5 & $7.5 \pm ^{0.2}_{0.3}$ & 5.6 & 64\% \\
B1 & 127 & $8.1 \pm ^{2.1}_{2.0}$ & 6.4 & 67\% & 124.5 & $7.3 \pm 1.0$ & 4.6  & 23\% \\
C1 & 123 & $8.2 \pm 1.7$ & 6.3 & 72\% & 124.5 & $6.6  \pm 2.3$ & 5.0 & 57\% \\ \hline
A2 & 125.5 & $7.7 \pm ^{2.4}_{3.6}$ & 7.3 & 91\% & 136 & $7.0 \pm ^{9.3}_{5.7}$ & 7.0 & 91\% \\
B2 & 125 & $8.6 \pm ^{2.6}_{2.7}$ & 7.0 & 87\% & 123.5 & $5.8 \pm ^{4.1}_{4.8}$ & 6.4 & 83\% \\
C2 & 125.5 & $6.2 \pm ^{4.2}_{6.4}$ & 6.1 & 79\% & 127 & $4.4 \pm ^{8.4}_{9.4}$ & 6.9 & 91\% \\
\hline

& \multicolumn{4}{c|}{Max. Northerm Hemisphere Ice} & \multicolumn{4}{c|}{Max. Southern Hemisphere Ice}\\ \hline
True & & 6.6  &  & &  & 1.1 & & \\ \hline
A1 & & $6.0 \pm 5.9$ & 1.3 & 69\% & & $2.3 \pm 5.6$ & 0.2 & 92\% \\
B1 & & $6.3 \pm 6.1$ & 1.6 & 70\% & & $2.4 \pm 5.7$ & 0.6 & 93\% \\
C1 & &$5.6 \pm 6.1$ & 1.7 & 72\% & & $2.7  \pm 5.7$ & 0.7 & 94\% \\ \hline
A2 & & $5.6 \pm 6.6$ & 2.6 & 78\% & & $1.7 \pm 5.8$ & 1.0 & 95\% \\
B2 & & $6.7 \pm 6.3$ & 2.2 & 77\% & & $2.2 \pm 5.8$ & 0.0 & 92\% \\
C2 & & $4.0 \pm ^{7.5}_{8.9}$ & 2.5 & 78\% & & $1.4 \pm 6.0$ & 0.6 & 93\% \\ \hline
\end{tabular}

67\% confidence intervals shown.
\label{table:validationstats_gsl}
\end{table*}

\subsection{Global analysis}

\begin{figure*}[tb] 
   \centering
   \includegraphics[width=6.5in]{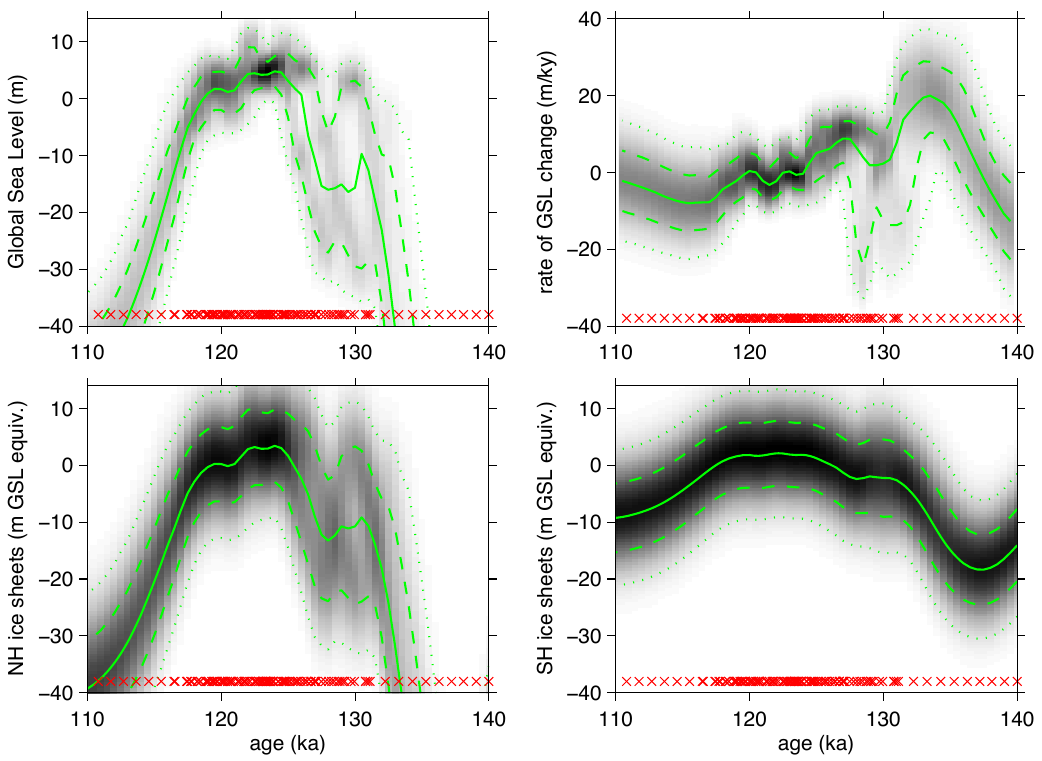} 
   \caption{Probability density plots of global sea level, 1000-year average global sea level rates, Northern Hemisphere ice volume and Southern Hemisphere ice volume projected from our full database. Dashed lines mark the 16th and 84th percentiles; dotted lines mark the 2.5th and 97.5th percentiles. Crosses at bottom mark the median posterior estimates of the sample ages.}
   \label{fig:slanalysis_full}
\end{figure*}

Applying our algorithm to the full data set of LIG sea level indicators (Tables \ref{table:slanalysis_gsl} and Figure \ref{fig:slanalysis_full}) reveals a peak median GSL of $4.8 \pm 2.7$ m (67\% confidence interval) centered at 124~ka. The 95\% and 80\% probability exceedance value for GSL are 5.8 and 6.8 m. This result is sensitive to the subset of the data examined (Figure \ref{fig:slanalysis}). Excluding the global $\delta^{18}$O curve but retaining the Red Sea curve yields a peak median GSL of $9.6 \pm 3.2$ m, consistent with some of the high values that characterize the Red Sea curve. (The Red Sea curve itself has a peak value of $12.4 \pm  3.0$ m ($1\sigma$).) The associated 95\% and 80\% probability exceedance values are 7.9 and 9.3 m. Excluding both the oxygen isotope curve and the Red Sea curve yields a peak median GSL of $5.7 \pm 4.3$ m and 95\% and 80\% probability exceedance values of 7.6 and 9.1 m. We therefore conclude that global sea level during the Last Interglacial indeed reached significantly higher levels than present, probably in the range of 6--9 m higher.

The 95\% and 80\% probability exceedance values for 1000-year average GSL rise rate during the interval when GSL was $\geq -10$ m range from 8.2 m/ky to 10.7 m/ky.  We emphasize that these values by no means exclude faster intervals of sea level rise lasting for a few centuries.

As expected from our validation analysis, the data is insufficient to make strong statements about the source of the meltwater that fed higher sea levels. They do, however, indicate that both the Northern Hemisphere ice sheets and the Antarctic ice sheets contributed. The analysis of the full data set indicates that the minimum NH ice volume was $3.4 \pm 6.4$ m GSL equivalent smaller than today and that the minimum SH ice volume was $2.1 \pm 5.7$ m smaller than today. The ranges defined by the 95\%  and 80\% probability exceedance values for each hemisphere are respectively 0.2--4.0 m and 0.3--3.5 m.

The two subset analyses indicate smaller ice volumes, consistent with the higher global sea levels they also indicate. The case without the global oxygen isotope curve indicates a minimum NH ice volume of $7.5 \pm 6.5$ m GSL equivalent and a minimum SH ice volume of $2.0 \pm 5.7$ m GSL equivalent, with ranges defined by the 95\% and 80\% probability exceedance values of 5.4--8.9 m and 2.3--5.3 m. The case without both the global oxygen isotope curve and the Red Sea curve indicates a minimum NH ice volume of $4.5 \pm 7.3$ m but is otherwise nearly identical to the case with the Red Sea: a minimum SH ice volume of $1.8 \pm 5.8$ m, and ranges of of 5.5--9.0 m and 2.5--5.5 m. 

\begin{table*}[tbp]
\renewcommand\arraystretch{1.3}
\textbf{\caption{Summary Statistics from Data Analysis}}
\scriptsize
\begin{tabular}{|l|rr|r|r|rr|r|r|}
\hline
& \multicolumn{4}{c|}{Max. Global Sea Level} & \multicolumn{4}{c|}{GSL Rate at $\geq$ -10 m}\\ \hline

& \multicolumn{2}{c|}{Median Projection} & \multicolumn{2}{c|}{Exceed. Level (m)} & \multicolumn{2}{c|}{Median Projection} & \multicolumn{2}{c|}{Exceed. Level (m/ky)}  \\
Case & Time (ka) & Max. (m) & 95\% & 80\% &  Time (ka) & Max. (m/ky) & 95\% & 80\%  \\
\hline
Full Data Set & 124.0  & $4.8 \pm 2.7$ & 5.8 & 6.8 & 126.0 & $6.2 \pm ^{5.6}_{6.5}$ & 9.0 & 10.4 \\
No $\delta^{18}$O & 123.5 & $9.6 \pm 3.2$ & 7.9 & 9.3 & 126.0 & $10.3 \pm ^{3.7}_{4.4}$ & 9.2 & 10.7 \\
No $\delta^{18}$O or RS & 129.5 & $5.7 \pm 4.3$ & 7.6 & 9.1 & 133.0 & $6.1 \pm ^{8.1}_{9.4}$ & 8.2 & 10.1 \\ 
\hline

& \multicolumn{4}{c|}{Max. Northern Hemisphere Ice} & \multicolumn{4}{c|}{Max. Southern Hemisphere Ice}\\ \hline

Full Data Set & & $3.4 \pm 6.4$ & 0.2 & 4.0 & & $2.1 \pm 5.7$ & 0.3 & 3.5  \\
No $\delta^{18}$O & & $7.5 \pm 6.5$ & 5.4 & 8.9 & & $2.0 \pm 5.7$ & 2.3 & 5.3 \\
No $\delta^{18}$O or RS & & $4.5 \pm 7.3$ & 5.5 & 9.0 & & $1.8 \pm 5.8$ & 2.5 & 5.5 \\
\hline

\end{tabular}

 67\% confidence intervals shown for the maximum of the median projections. 95\% and 80\% probability exceedance levels, which we employ as estimates of the maxima, are levels exceeded in 95\% and 80\% of all histories sampled from the estimated posterior distribution, respectively.
\label{table:slanalysis_gsl}
\end{table*}

\subsection{Cross-validation analysis of researcher bias}

One challenge in integrating local sea level records produced by many researchers is researcher bias. We therefore conducted a cross-validation study by classifying the data points by research group and repeating the global analysis on subsets of data in which we exclude all results from one research group. To compare the results of analyses on different subsets, we calculated the Mahalanobis distances between them in terms of global sea level, as described in section \ref{sect:summarystats}. As controls, we calculated the distance between our full analysis and nine individual parallel runs of the Monte Carlo simulation on the full data set.

When measured in terms of global sea level, only one subset was more different from the full analysis than all nine control runs: the subset excluding the global $\delta^{18}$O curve. Given the number of points in this curve, this result should be expected \emph{a priori}.

\subsection{Outlier Analysis}

To search for outliers, we estimated the posterior probabilities for each of our sea level measurements and age measurements given the distribution at each point for sea level and age projected by our statistical model. We performed this analysis both upon the results produced from the full data sets and the results produced from the subset excluding the $\delta^{18}$O curve and the Red Sea curve.

No data point was a strong outlier, but three sites generated sea level measurement probabilities between 0.15 and 0.33. Global sea level at 132, 131, and 129~ka was moderately higher than the values inferred from the global $\delta^{18}$O curve; inferred sea levels of  $-58 \pm 15$ m, $-35 \pm 10$ m and $-40 \pm 11$ m (1$\sigma$) gave rise to projected GSL of $-28 \pm 15$ m, $-13 \pm ^{12}_{16}$ m and $-27 \pm 7$ m. These estimates suggest a relationship between $\delta^{18}O$ and sea level of about -30 m/$\permil$, a somewhat surprising result because such low slopes are more characteristic of glaciations than deglaciations \citep{Waelbroeck2002a}.

 At Kahe Beach State Park, Oahu, Hawai`i, \citep{Hearty2007b} describe a marine conglomerate at 12 m above present sea level. Corrected for uplift of Oahu, this suggests a paleo-sea level of at least $9.6 \pm 1.3 (1\sigma)$ m. Our model instead assigns a sea level of $7.5 \pm 1.7$ m. When the $\delta^{18}$O and Red Sea curves are not included, it assigns a sea level of $8.0 \pm ^{2.8}_{2.4}$ m. These results suggest that the sea level indicator should be reexamined.
 
 Finally, our model identifies as an outlier early Weichselian (post-Eemian) lacustrine sediment from a boring in the North Sea \citep{Zagwijn1983a}. The sediment indicates freshwater conditions at a relative sea level of about -40 m, which we adjust to $-23 \pm 3$ m based upon the subsidence estimates of \citet{Kooi1998a}. The model, however, places sea level at $-18.5 \pm ^{6.5}_{8.9}$ m.  In the absence of the $\delta^{18}O$ and Red Sea curves, it places sea level at $-10.4 \pm 4.4$ m. In the absence of  these curves, the model also identifies as an outlier early Eemian lacustrine sediment in the same core. These results suggest that the North Sea in the region of this boring is subsiding faster than the \citet{Kooi1998a} estimates.

\section{Discussion}

\subsection{Comparison to past estimates of Last Interglacial sea level}

The Fourth Assessment Report of the IPCC \citep{Jansen2007a} estimates that LIG sea level reached values of 4--6 m above present. Their estimate is qualitatively similar to the result of our analysis, but ascribes an excessive degree of precision and may be somewhat low. Our analysis suggest that sea level peaked during the Last Interglacial between about 5 and 9 meters above present. A more thorough analysis of sea level records, in other words, seems to permit even higher sea level than the IPCC's figure. 

The IPCC summary estimate, like other similar estimates \citep[e.g.,][]{Overpeck2006a}, is based upon examining a few key coral reef terrace localities. The IPCC highlights Hawaii and Bermuda \citep{Muhs2002a}; \citet{Overpeck2006a} also highlight the Bahamas, Western Australia, and the Seychelles Islands. All these localities are basically tectonically stable and experience slow thermal subsidence. If one had to draw conclusions about global sea level from a small number of local sea level measurements, these would be reasonable sites at which to look.

Other commonly considered localities, such as Barbados \citep[e.g.,][]{Schellmann2004a} and the Huon Peninsula \citep{Esat1999a}, are rapidly uplifting localities. These sites have advantages as relative sea level recorders, most notably that terraces recording sea levels below present are readily accessible. Assuming these sites have experienced a steady rate of uplift, they can help uncover sea level variations over fairly short timescales. However, they are poor sites from which to draw conclusions about absolute sea levels, as recovering this information requires a precise estimate of uplift rate.

To our knowledge, no previous author has accounted for the effects of glacial isostatic adjustment (GIA) in drawing conclusions about global sea level and ice volume from Last Interglacial sea level records. As \citet{Lambeck1992a} demonstrated, understanding the influence of these effects is critical. Without this understanding, local sea level highstands could be falsely interpreted as reflecting global highstands. In the mid-to-late Holocene, for instance, GIA has produced local highstands in far-field equatorial islands of about 1--3 m above present levels \citep{Mitrovica1991a}; looking only at these sites in isolation, one might falsely infer that global sea level was higher and global ice volume significantly smaller in the mid-Holocene than at present. Our statistical model uses the covariance between local and global sea level to correct for these effects; our results indicate that the apparent high Last Interglacial global sea levels are real. 

\subsection{``Fingerprinting'' analysis of meltwater sources}

Just as our model can predict global sea level from local sea level measurements, it can also predict changes in ice sheet volumes. However, as the validation analysis showed, the current database combined with the simple GIA model we employ cannot predict these changes with great precision (Figure \ref{fig:slanalysis_full}c,d).

The analysis does suggest that it is likely both that melting of northern hemisphere ice sheets and also melting of the Antarctic ice sheet contributed to higher sea levels. More near-field sea level measurements would help increase the precision of these estimates; however, incorporating near-field measurements into the analysis requires a more sophisticated physical model. Our model correctly accounts for neither isostatic uplift nor flexure, both of which are important in interpreting near-field sites. Indeed, even though we include some near-field data from Svalbard and Greenland in our database, we could not incorporate these points into our analysis.

However, our approach could readily be adapted to incorporate a sophisticated glacial isostatic adjustment model like that of \citet{Mitrovica2003a}. Such a model could be used to generate a spatial and temporal covariance function in place of our simple model. To do so would require a method for randomly generating hundreds of plausible ice sheet histories extending from before the Last Interglacial to today and modeling the full gravitational, elastic, and isostatic effects of these changing ice sheet volumes on local sea level around the globe. Such an effort would be computational much more intensive than employing our simple model, which takes less than fifteen minutes on a desktop computer to run 300 simulations. Nonetheless, it is quite feasible.

\subsection{The need for more data and opportunities for future research}

\begin{figure}[tb] 
   \centering 
  \includegraphics[width=3.25in]{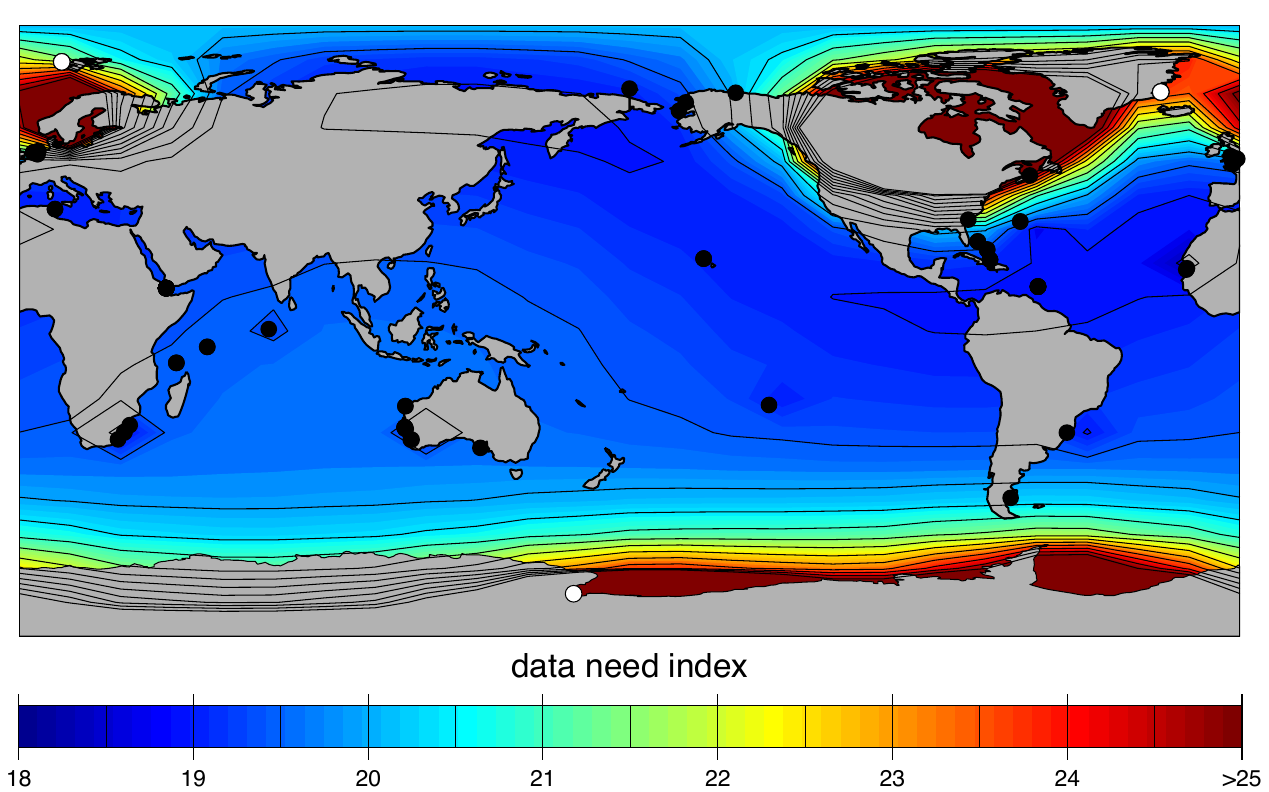} 
   \caption{Map of the data need index. We calculate this index by averaging the ratio of the posterior standard deviation to the prior standard deviation over the time period between 118 and 130~ka. Units are percentages. Dots indicate the positions of observations. Open dots are observations not included in the analysis.}
   \label{fig:dataneedindex}
\end{figure}

Mapping the uncertainty of local sea level projections highlights the regions whence more data is most strongly needed. The ``Data Need Index'' (Figure \ref{fig:dataneedindex}) is the mean of the ratio of the posterior standard deviation to the prior standard deviation over the time period between 118 and 130~ka. The greatest need is in the near-field of the major ice sheets; as noted above, the need in these regions is to both expand the database and employ a more sophisticated GIA model in our algorithm. The second greatest need for data is in the southern Indian Ocean, the regional antipodal to the Laurentide Ice Sheet, and thus more sensitive than the rest of the intermediate and far-field to GIA effects associated with Laurentide melting. Studies in areas such as Madagascar, Mauritius, and the islands of the French Southern and Antarctic Lands would be helpful in this regard.

In compiling the LIG sea level database, we also found a number of region where sea level indicators require further investigation. For instance, although Britain is on a tectonically stable passive margin, erosional terraces appear to get progressively older with increasing elevation. \citet{Westaway2006a} estimated Pleistocene uplift rates in the vicinity of the Solent river system range of $\sim 10$ m/ky. The causes of this uplift are uncertain, but might be linked to isostatic effects caused by erosional unroofing and the transport of sediment from continent to slope.  A simple isostatic calculation indicates this method requires the removal of $\sim$50 m of sediment per 100 ky. \citet{Clayton1996a} estimates that an average thickness of $\sim$145 m of sediment was removed from the land of the British Isles to the continental shelf during the last  glaciation; this removal could therefore be a potential cause. Because the British Isles are in a crucial region to look for the fingerprint of Greenland melting, a better understanding of regional uplift would be extremely helpful 

\citet{Braithwaite1984a} described numerous terraces in the coastal limestone of Kenya which range in elevation from -35 m to +20 m but lack good age constraints. These represent ready targets for modern dating techniques.

\subsection{Rates of sea level change}
we expect that ice sheet dynamics during these intervals will more closely resemble future ice sheet dynamics than will the behavior of ice sheets during intervals of lower GSL.

Our results suggest that, during the interval of the Last Interglacial when sea level was $-10$ m or higher, the rate of sea level rise, averaged over one thousand years, reached values of at least about 8--11 m/ky. Within the resolution of our data, we cannot confidently resolve rates of sea level change over shorter periods of time. These values are consistent with \citet{Carlson2008a}'s estimates of the rate of the contribution of the Laurentide Ice Sheet meltwater to global sea level during the early Holocene; they estimate a Laurentide contribution of about 7 m/ky during the period when global sea level climbed above $-10$ m.

Ice volume during the late deglacial rise at the start of the Last Interglacial was only slightly larger than at present. The Laurentide Ice Sheet would have been a shrunken remnant of its once extensive mass -- or, perhaps two small remanents, one over Quebec and Labrador and one over eastern Nunavut and Baffin Island, as in the late early Holocene \citep{Carlson2008a}. It was within a factor of two in size of the present Greenland Ice Sheet, and its dynamics may therefore have been analogous to those of the Greenland Ice Sheet. Given a sufficient forcing, the results from the Last Interglacial suggest that the present ice sheets could sustain a rate of global sea level rise of about 80--110 cm/century for several centuries, with these rates potentially spiking to higher values for shorter periods.

\section{Conclusion}

Contrary to an analogy commonly taught in introductory classes, adding water from melting land ice to the ocean is not like pouring water into a bathtub. Many factors other than the changing volume of water in the ocean modulate the influence of melting ice sheets on local sea level. These factor include: the effects of the distribution of ice, water, and sediment on the geoid; lithospheric flexure; isostatic adjustment; and tectonic uplift and subsidence, as well as dynamic effects, which are of lesser concern on multi-century timescales.

Consequently, global sea level and global ice volume cannot be accurately inferred simply by examining local sea level at one or two localities, yet this is the path most commonly taken when discussing the Last Interglacial. Our approach, which combines an extensive database with a new statistical algorithm for analyzing quantitative paleoenvironmental data with both interpretive and geochronological errors, offers better control. The results of our analysis support the common hypothesis that Last Interglacial global sea level was higher than present. We find that peak GSL was probably 6 to 9 m higher than present.

The Last Interglacial was only slightly warmer than present, with polar temperatures similar to those expected under a low-end, $\sim2^\circ$C warming scenario. Nonetheless, it appears to have been associated with substantially smaller ice sheets than exist at present. Our results indicate that both the Greenland Ice Sheet and the Antarctic Ice Sheets were at least smaller than today. Global sea levels five meters higher than present could have been produced by a significantly smaller Greenland Ice Sheet, a significantly smaller Antarctic Ice Sheet, or both. Global sea levels nine meters higher than present would have required significantly smaller ice sheets in both hemisphere, including nearly complete melting of either the Greenland and the West Antarctic Ice Sheets. This paleoclimatic constraint emphasizes the vulnerability of ice sheets to even relatively low levels of sustained global warming. 


{\small \section*{\small Acknowledgements}

Computing resources were substantially provided by the TIGRESS high performance computer center at Princeton University, which is jointly supported by the Princeton Institute for Computational Science and Engineering and the Princeton University Office of Information Technology. REK was funded by a STEP Postdoctoral Fellowship. }


{\small
\bibliographystyle{REK-elsart-harv}
\bibliography{/Users/bob/Literature/BibdeskRefs-RKopp}
}

{\appendix
\section{Database}

Our database of Last Interglacial sea level indicators is based on a literature search for indicators with best estimates of ages between 140 and 110~ka. We characterized each candidate indicator by several quantitative parameters:
\begin{itemize}
\item Location (latitude and longitude)
\item Age estimate
\item Altitude of indicator
\item Indicative meaning (i.e., the relationship between indicator and sea level)
\item Estimate of tectonic uplift or subsidence rate
\item Stratigraphic order, when more than one point comes from the same site; where available, we also include estimates of the relative ages of points at the same site
\end{itemize}

The database is recorded in a spreadsheet that accompanies this supplemental material. Three of the sites are re-analyses of data available elsewhere that require special explication: the sea level curve derived from the \citet{Lisiecki2005a} global oxygen isotope curve, the re-aligned Red Sea sea level curve of \citet{Rohling2008a}, and a subsidence-corrected Dutch sea level curve based on \citet{Zagwijn1983a}.

\subsection{Global oxygen isotope stack}

By making assumptions about the isotopic composition of ice sheets and a paleotemperature constraint, it is possible to estimate ice volume and thus global sea level from carbonate $\delta^{18}$O. These records most commonly come from the shells of foraminifera, as is the case for the global oxygen isotope stack of  \citet{Lisiecki2005a} (LR04). Independent paleotemperature estimates can be derived from the Mg/Ca ratio of foraminifera \citep[e.g., ][]{Lea2002a}, alkenone composition \citep[e.g., ][]{Haywood2005a}, or clumped isotopes \citep{Ghosh2006a}. Alternatively, a sea level record can be derived by modeling or making assumptions about the relationship between ice sheet volume, changes in ice sheet isotopic composition, and temperature \citep[e.g., ][]{Bintanja2005a}.

The simplest way to convert these oxygen isotopic measurements into a measure of sea level is to calibrate the curve against modern and Last Glacial Maximum values. By definition, modern global sea level is zero meters; with reference to the VPDB oxygen isotope standard, the modern benthic foraminifera oxygen isotopic composition from the LR04 stack is $3.23 \pm 0.06 \permil$. The Last Glacial Maximum value (at 18~ka) is $5.08 \pm 0.06 \permil$, while sea level was $\sim-125$ m \citep[e.g.,]{Peltier2004a}. From this result,  one would derive a conversion factor of $\sim-67.6 \pm 7.2$ m/$\permil$. Based on comparison of coral records with oxygen isotopes, however, \citet{Waelbroeck2002a} observe that this conversion is not always linear. During the onset of glaciations, oxygen isotopes tend to respond more quickly than sea level, while during terminal deglaciations, sea level responds more quickly. These changes reflect changes in both ice sheet composition and the relationship between temperature and ice sheet melt. Near modern sea levels, North Atlantic records indicate a relationship during glaciation of $\sim30$ m/$\permil$, while during glacial terminations, the relationship is $\sim90$ m/$\permil$. We therefore use a relationship of $60 \pm 30$ m/$\permil$ in our calculations. When this uncertainty is combined with measurement uncertainty, it yield a peak Last Interglacial value of $7.8 \pm 8.0$ m (from an oxygen isotopic composition at 123~ka of $3.10 \pm 0.10 \permil$. 

The absolute ages of the LR04 stack between 22 and 120~ka are derived by alignment against the oxygen isotopic record of the  MD95-2042 core, from the Iberian margin \citep{Shackleton2000b}. This core is, in turn, aligned between 0 and 67~ka against the oxygen isotopic record of the GRIP ice core \citep{Johnsen1992a}.  Termination II is assigned to start after $\sim$135~ka based upon U-Th dating of corals terraces from Papua New Guinea \citep{Stein1993a}.  While this age model is not necessarily superior to alternative age models (for instance, that employed by \citet{Rohling2008a}), we have aligned the other quasi-continuous records against it so as to provide a common reference frame.

\begin{figure}[tb] 
   \centering
   \includegraphics[width=3.25in]{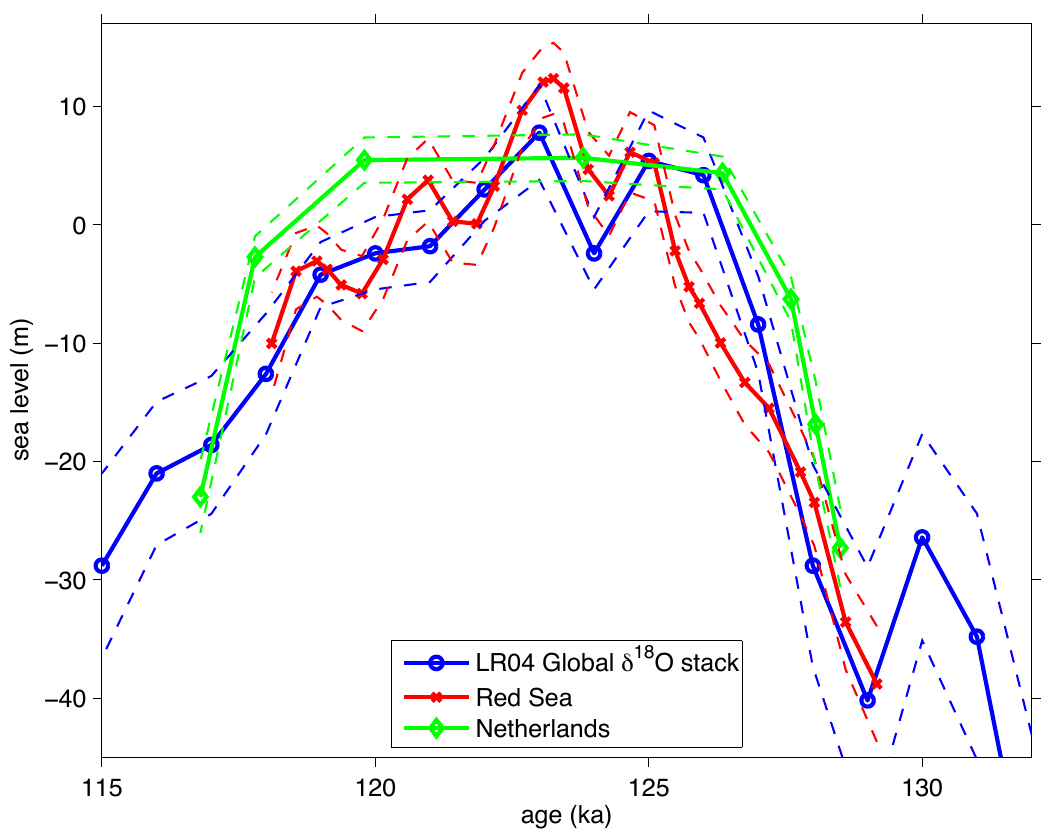} 
   \caption{The LR04 global sea level curve and local sea level curves from the Red Sea and the Netherlands. Dashed lines show 1$\sigma$ confidence intervals in sea level. The initial best alignment of the three curves is shown.}
   \label{fig:quasicontinuous}
\end{figure}

\subsection{Red Sea}

The Red Sea record is a planktonic foraminiferal oxygen isotope record that takes advantage of the particular hydrology of the Red Sea \citep{Siddall2003a} and is therefore essentially a local record of sea level at the Strait of Bab el Mandab. The oxygen isotopic composition of Red Sea water is controlled primarily by evaporation. Water exchange occurs between the Red Sea and the Indian Ocean occurs through the strait; when sea level is lower, water exchange decreases, which increases the residence time of water in the Red Sea and thus yields heavier oxygen isotope values. This greatly magnifies the isotopic effects of sea level change. The difference between the modern and the Last Glacial Maximum in the Red Sea is nearly 6$\permil$, whereas in the open ocean the difference is approximately 1.8$\permil$.
				
Using a hydrological model, \citet{Rohling2008a} constructed a sea level record with a raw 1$\sigma$ precision of 6 m for the Last Interglacial from two Red Sea cores sampled for oxygen isotopes at 10 cm resolution. They aligned their record temporally with the record derived from U/Th-dated Barbados coral data \citep{Thompson2005a}; in this age model, their record has a temporal resolution of 200--400 years. It indicates that local sea level rose to at least 6 $\pm$ 3.5 m, and perhaps as high as 11 m, during the peak interglacial.
	
We have for consistency realigned the \citet{Rohling2008a} against the age model for the global oxygen isotope stack of \citet{Lisiecki2005a}, which is based primarily on alignment against the GRIP ice core. This realignment required shifting the curve earlier by 2.4~ka and expanding the duration between measurements 1.2 times. We include in our database the re-aligned sea level curve derived from the KL11 core, while Rohling et al. (2008) provide a higher resolution record than the KL09 core.

\subsection{Netherlands}

The Dutch Eemian sea level record of \citet{Zagwijn1983a} is based on sedimentological and micropaleontological data from numerous cores through the Amsterdam and Amersfoort basins, as well as cores along the Noord-Holland coast, in Friesland, and in the North Sea. Sea level indicators in these cores are provided by facies transitions representing, for example, the infiltration of marine water into a freshwater lake or the maximum elevation of clays deposited in a salt-marsh environment. Relative age constraints are provided by characteristic Eemian pollen zones, many of which have durations established to fairly high precision based upon the counting of varves in an annually-layered lacustrine diatomite in northwestern Germany \citep{Zagwijn1996a}.  We place peak sea level in the middle third of zone E5 based upon the position of the maximum flooding interval within the more recent Amsterdam-Terminal borehole \citep{Leeuwen2000a}. We estimate absolute ages from these relative ages by aligning the sea level curve against the \citet{Lisiecki2005a} global oxygen isotope stack.

Zagwijn reported sea level estimates without correction for long-term isostasy, compaction, or tectonics. To correct for these factors, we use the backstripping-derived Quaternary rate estimates of \citet{Kooi1998a}. These vary considerably across the Netherlands and the North Sea, ranging from about 12 cm/ky in Amersfoort to about 18 cm/ky in Petten. Thus adjusted, Zagwijn's data indicate that a maximum local sea level of about $5 \pm 2$ m was attained in the Netherlands for much of the Last Interglacial.

\begin{figure}[tb] 
   \centering
   \includegraphics[width=3.25in]{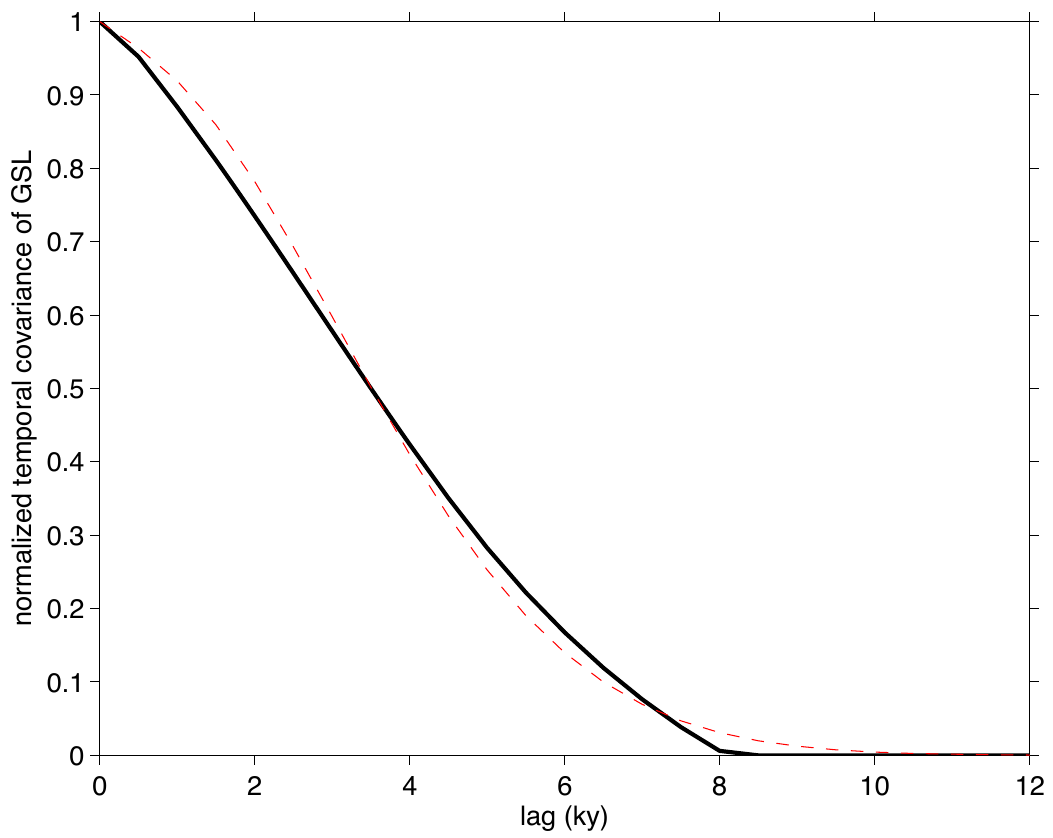} 
   \caption{The temporal covariance function of global sea level (black) and the Gaussian fit used to approximate it in our model (red).}
   \label{fig:temporalcov}
\end{figure}

\begin{figure*}[tb] 
   \centering
   \includegraphics[width=6.5in]{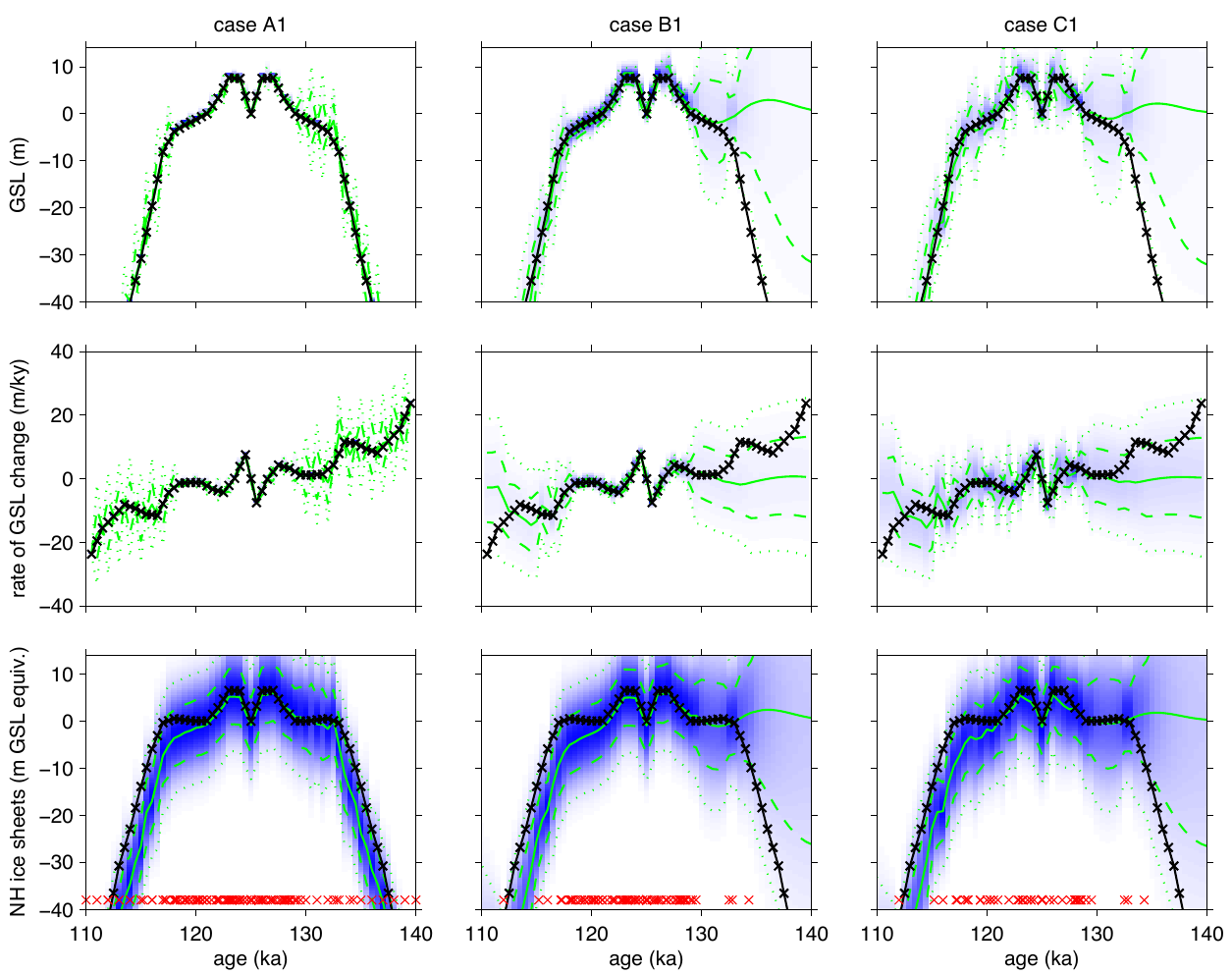} 
   \caption{Probability density plots (blue) of global sea level (top row), 1000-year average global sea level rates (middle row) and Northern Hemisphere ice volume (bottom row) as recovered by sampling a synthetic data set (black).  The synthetic data was sampled with no temporal errors and 10 cm elevation errors. Case A1 includes all data, case B1 excludes the global $\delta^{18}O$ curve, and case C1 excludes both the global $\delta^{18}O$ and the Red Sea data. Dashed lines mark the 16th and 84th percentiles; dotted lines mark the 2.5th and 97.5th percentiles. Crosses at bottom mark the median posterior estimates of the sample ages.}
   \label{fig:validation_noerrors}
\end{figure*}

\begin{figure*}[tb] 
   \centering
   \includegraphics[width=6.5in]{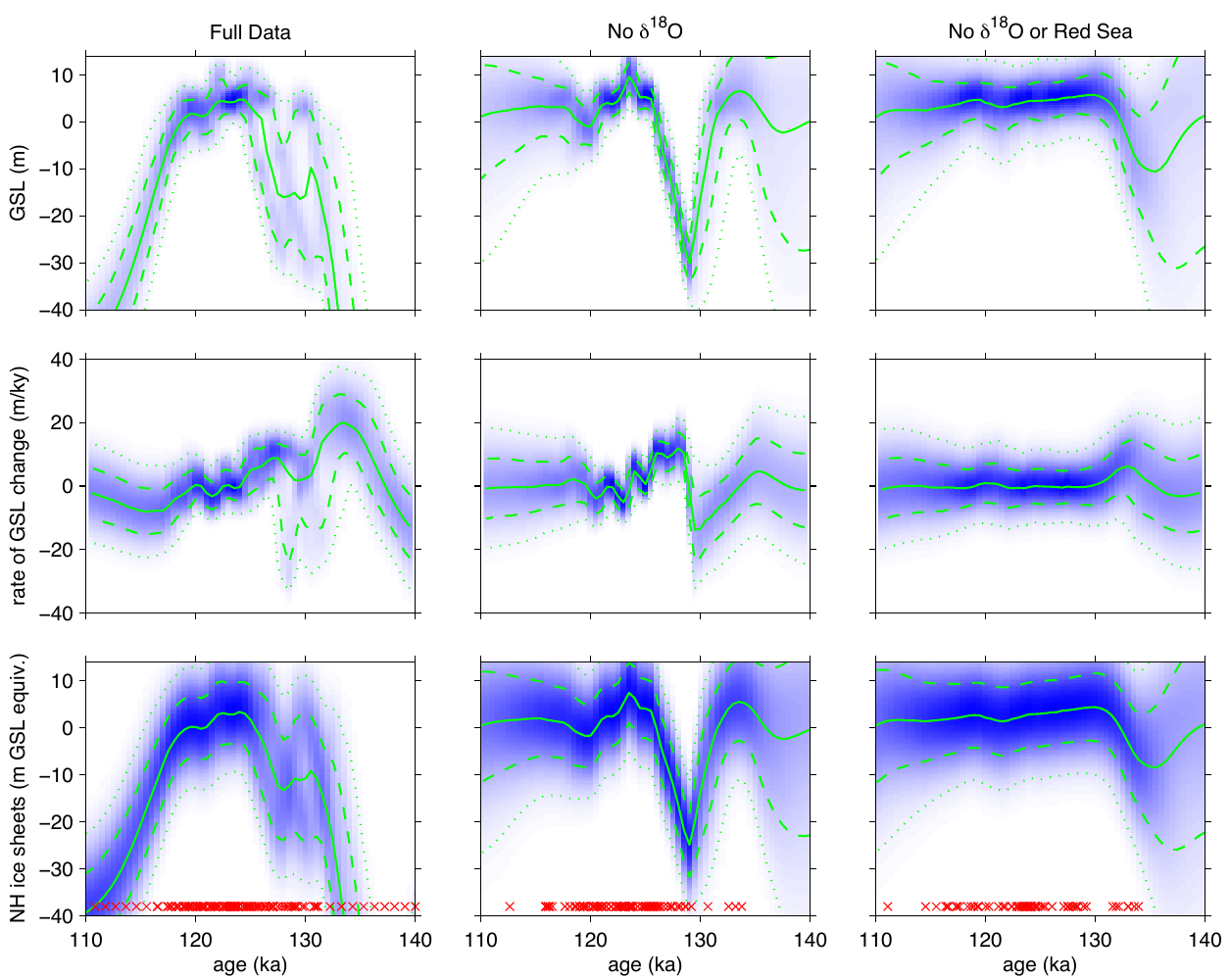} 
   \caption{Probability density plots (blue) of global sea level (top row), 1000-year average global sea level rates (middle row) and Northern Hemisphere ice volume (bottom row) projected from our full database (left column), our database excluding the global $\delta^{18}O$, and our database excluding both the global $\delta^{18}O$ and the Red Sea data. Dashed lines mark the 16th and 84th percentiles; dotted lines mark the 2.5th and 97.5th percentiles. Crosses at bottom mark the median posterior estimates of the sample ages.}
   \label{fig:slanalysis}
\end{figure*}
}

 \end{document}